\documentclass[5p,twocolumn,times]{elsarticle}
\usepackage{cmap}
\usepackage[T1]{fontenc}
\usepackage[utf8]{inputenc}
\usepackage{amsmath,amssymb}
\usepackage{graphicx}
\usepackage{subcaption}
\usepackage{booktabs}
\usepackage{array}
\usepackage{xcolor}
\usepackage{listings}
\usepackage{enumitem}
\usepackage{url}
\usepackage[hidelinks]{hyperref}
\usepackage{stfloats}
\usepackage{float}
\usepackage[none]{hyphenat}

\newcommand{\code}[1]{\texttt{#1}}

\newcommand{\cfg}[1]{\path{#1}}
\newcommand{\grainsmith}{\textsc{grainsmith}}
\newcommand{\sgnum}{230}

\lstdefinestyle{yaml}{
	basicstyle=\ttfamily\footnotesize,
	keywordstyle=\color{blue!60!black},
	commentstyle=\color{gray},
	breaklines=true,
	frame=single,
	showstringspaces=false,
	columns=fullflexible
}

\newcounter{bla}

\makeatletter
\def\ps@pprintTitle{%
  \let\@oddhead\@empty
  \let\@evenhead\@empty
  \def\@oddfoot{\hfil\thepage\hfil}%
  \let\@evenfoot\@oddfoot
}
\makeatother

\begin{document}
	
	\begin{frontmatter}
		
        \title{\grainsmith{}: A Generator of Polycrystalline Models for Atomistic Simulations with Statistical and Grain-Boundary Morphology Control}
		
		\author[firat]{Oguzhan Orhan\corref{cor1}}
		\ead{oguzhan.orhan@firat.edu.tr}
		\cortext[cor1]{Corresponding author. ORCID 0000-0003-2049-053X.}
		
		\author[firat]{Soner \"Ozgen\fnref{orcidsoner}}
		\ead{sozgen@firat.edu.tr}
		\fntext[orcidsoner]{ORCID 0000-0003-4292-9187.}
		
		\address[firat]{Department of Physics, F\i rat University, Elaz\i\u{g}, T\"urkiye}
		
		\begin{abstract}
			Atomistic studies of grain-boundary engineering, grain-size effects and dopant enrichment require reproducible models
			with prescribed microstructural features. We present \grainsmith{}, an open-source Python package for generating
			statistically controlled polycrystalline models with selectable grain-boundary morphologies for molecular dynamics and
			subsequent relaxation. Within supported feature combinations, a single configuration specifies grain-size and volume
			distributions, crystallographic texture, boundary-area-weighted disorientation-angle distributions, phase composition and grain-boundary dopant placement. Crystal construction supports \sgnum{} crystallographic space groups. Periodic Voronoi and volume-targeted Laguerre tessellations provide planar boundaries, while distinct geometry backends generate smoothly curved and band-limited self-affine boundaries. A registry of twenty-six checks assesses applicable inputs, construction properties and outputs. Each run exports LAMMPS data and Extended XYZ files together with structural and statistical descriptors and machine-readable provenance. For a fixed software version and computational environment, the configuration and random seed determine byte-reproducible
			atomic configurations and scientific data across supported worker counts. By combining statistical specification,
			boundary-morphology control and reproducible atomistic output, \grainsmith{} supports systematic studies of microstructural effects and quantitative comparisons across generated models.

		\end{abstract}
		
		
	\end{frontmatter}
	
	\section{Introduction}
	\label{sec:motivation}
	Many material properties studied using molecular dynamics (MD) are sensitive to
	microstructure: mechanical response depends on grain size and crystallographic
	texture, while transport, fracture, corrosion, and phase transformation can
	depend strongly on grain-boundary (GB) and interphase networks. Consequently,
	atomistic simulations of polycrystalline materials often require
	microstructural statistics to be specified before atomistic construction rather
	than characterised only after a structure has been generated. Reproducible
	regeneration of such structures is also important when microstructure
	statistics are used as controlled variables in a simulation study.
	
	Several established tools address different parts of this design space.
	Atomistic generators such as Atomsk~\cite{hirel2015atomsk} and
	PolyPal~\cite{shin2025polypal} focus on direct construction of atomistic
	polycrystals, with different emphases on flexibility and large-scale parallel
	generation. Atomsk constructs polycrystals by expanding a supplied atomic seed
	into a Voronoi tessellation and supports both randomly generated and explicitly
	specified grain positions and crystallographic orientations
	~\cite{hirel2015atomsk}. PolyPal extends atomistic polycrystal generation to
	large parallel systems, with explicit control over seed placement and
	grain-scale statistics together with support for multiphase structures and
	solute incorporation~\cite{shin2025polypal}. Other tools emphasize
	microstructure generation and statistical control at the mesoscopic or voxel
	level. Neper generates three-dimensional polycrystalline microstructures and
	can fit Laguerre tessellations to prescribed statistical cell properties,
	including grain-size and shape distributions
	~\cite{quey2011neper,quey2018laguerre}. DREAM.3D provides voxel-based synthesis
	and analysis of statistically specified microstructures, including
	crystallographic texture and other microstructural descriptors
	~\cite{groeber2014dream3d}. Kanapy provides synthetic polycrystalline
	microstructures with crystallographic texture reconstruction and orientation
	assignment designed to reproduce disorientation-angle statistics, including
	boundary-size weighting and, optionally, grain-volume weighting of the
	represented texture~\cite{prasad2019kanapy,biswas2020kanapytexture}. PyAPD
	provides GPU-accelerated anisotropic power diagrams fitted through
	optimal-transport methods, enabling statistically controlled polycrystalline
	geometries with explicit control of individual grain volumes
	~\cite{buze2024pyapd}. Dedicated bicrystal-oriented tools such as
	aimsgb~\cite{cheng2018aimsgb} and GB\_code~\cite{hadian2018gbcode} focus on
	constructing individual grain boundaries rather than statistically controlled
	three-dimensional grain-boundary networks. 
	
	These capabilities are complementary rather than mutually exclusive.
	The integration problem addressed here is to combine compatible
	microstructural controls within a reproducible atomistic workflow,
	with quantitative checks on the construction geometry and
	resulting atomic structure. In particular,
	users may need to control grain volumes or grain-size distributions,
	crystallographic texture, boundary-area-weighted disorientation-angle
	statistics, boundary morphology, phase fractions, and grain-boundary dopant
	enrichment while retaining a reproducible mapping from the input configuration
	to the final atomistic model. Existing tools provide important subsets of
	these capabilities, but they differ substantially in their representation
	level, construction mechanism, and scope of statistical control.
	
	\grainsmith{} addresses this integration problem by providing an atomistic
	construction pipeline in which these microstructural specifications are
	explicit inputs and the corresponding realised properties are re-measured
	during construction. The contribution is therefore not the invention of the
	individual underlying algorithms: periodic Laguerre tessellations fitted by
	semi-discrete optimal transport (SDOT)~\cite{bourne2020laguerre,bourne2023sdot},
	anisotropic power-diagram constructions~\cite{buze2024pyapd},
	orientation-assignment and boundary-disorientation optimisation strategies~\cite{prasad2019kanapy,biswas2020kanapytexture},
	band-limited self-affine surface generation~\cite{eder2017abrasion}, and
	crystallographic symmetry analysis~\cite{togo2018spglib} all have established
	precedents. Rather, \grainsmith{} integrates these approaches into a common
	atomistic workflow and adds a construction-time validation discipline that
	distinguishes exact construction identities, quantitative targets, diagnostic
	quantities, and known method limitations.
	
	The resulting generator provides seven complementary capabilities:
	\begin{enumerate}[label=(\roman*)]
		\item grain-volume and grain-size control through periodic Voronoi and
		semi-discrete-optimal-transport-fitted Laguerre tessellations, with the
		volume fit evaluated using an analytic SDOT Jacobian on explicitly
		reconstructed polyhedral cells;
		\item crystallographic texture construction together with
		boundary-area-weighted disorientation-angle shaping under an explicit
		volume-weighted ODF-drift constraint;
		\item curved and band-limited self-affine grain-boundary morphologies with
		independent Hurst back-estimation;
		\item multiphase construction with targeted and re-measured phase fractions;
		\item space-group-aware crystal construction supporting all 230
		three-dimensional crystallographic space groups through
		Wyckoff-orbit expansion and an \texttt{spglib} round-trip verification;
		\item deterministic regeneration, construction-time quality-assurance
		records, and machine-readable provenance; and
		\item deterministic grain-boundary-targeted dopant placement in
		substitutional and interstitial modes.
	\end{enumerate}
	
	A central design principle of \grainsmith{} is that statistical controls are
	reported together with their realisability limits and remain explicit throughout
	the generation workflow. The orientation-assignment annealer permutes a fixed
	set of grain orientations and therefore shapes, rather than guarantees, a target
	disorientation-angle distribution. Likewise, volume-weighted texture may drift
	when orientations are exchanged between grains of unequal volume, and this
	drift is explicitly re-measured. For self-affine boundaries, the prescribed
	Hurst exponent characterises a finite-band random field and is therefore
	distinguished from boundary-level roughness measured on the resulting geometry.
	Dopant enrichment is similarly defined by a geometric placement rule rather
	than predicted from thermodynamic segregation. These distinctions form part of
	the software contract rather than post hoc qualifications.
	
	This explicit treatment of statistical controls is complemented by an
	interoperable path from mesoscopic to atomistic representations. External voxel
	label fields can be imported and atomised, enabling microstructures generated
	or processed in DREAM.3D and other voxel-based workflows to enter the same
	atomistic pipeline. In this way, \grainsmith{} complements existing
	microstructure-generation ecosystems while extending statistically specified
	microstructures to atomistic starting configurations for molecular-dynamics
	studies.

	% =====================================================================
	\section{Implementation}
	\label{sec:architecture}
	
	This section describes the software itself: its pipeline architecture, parallel
	execution model, determinism contract, and the interfaces and outputs produced
	by each run. The geometric, crystallographic and statistical foundations of each
	pipeline stage are presented in Sec.~\ref{sec:theory}.
	
	As part of the research process, the authors used Claude Code (Anthropic),
	with the Claude Fable~5 and Claude Opus~5 models, as coding and debugging assistants
	during the development of \grainsmith{}. All such contributions were reviewed
	by the authors, and the behaviour of the resulting code is underwritten by the
	construction-time gate registry of Sec.~\ref{sec:gates} and the validation of
	Sec.~\ref{sec:validation}, rather than by its authorship.
	
	\subsection{Architecture and pipeline}
\begin{figure*}[!t]
		\centering
		\includegraphics[width=0.70\textwidth]{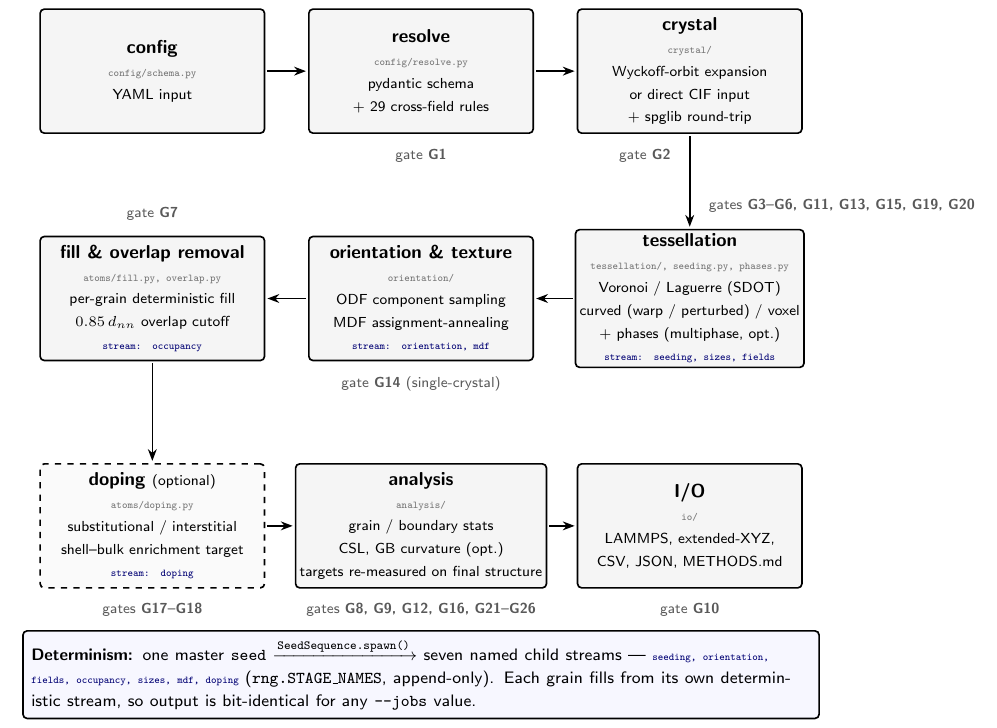}
		\caption{Software architecture and workflow of \grainsmith{}. Boxes are the
			pipeline stages (with their source subpackages); solid arrows indicate the
			one-way data flow from YAML configuration to the
			LAMMPS/extended-XYZ/CSV/JSON outputs. The crystal stage accepts either a
			space-group/Wyckoff specification or a CIF file, with symmetry
			re-detected by spglib on both routes. Quality-assurance gates
			(G1--G26, \ref{app:gates}) are annotated at the stage where each
			is evaluated and recorded, while coloured \code{stream:} tags mark where
			each of the seven named RNG child streams is consumed
			(Sec.~\ref{sec:determinism}). The dashed box denotes the optional doping
			stage, active only for doped single-phase configurations.}
		\label{fig:pipeline}
	\end{figure*}

	\grainsmith{} is organised as a one-way pipeline (Fig.~\ref{fig:pipeline}),
	orchestrated by \code{pipeline.run}:
	\code{config} $\rightarrow$ \code{resolve} $\rightarrow$ \code{crystal}
	$\rightarrow$ \code{tessellation} $\rightarrow$
	\code{orientation/texture} $\rightarrow$ \code{fill} $\rightarrow$
	\code{doping} (optional) $\rightarrow$ \code{analysis} $\rightarrow$
	\code{I/O}. The \code{resolve} stage uses a Pydantic schema with twenty-nine
	numbered cross-field rules that reject inconsistent configurations before
	execution; these are distinct from, and evaluated before, the twenty-six
	construction-time QA checks of Sec.~\ref{sec:gates}. The engine is deliberately
	dependency-light (NumPy, SciPy, spglib, PyYAML, and Pydantic), while the
	direct-CIF input path adds \code{ase} as an optional extra, making the workflow
	suitable for headless HPC environments.
	
	\subsection{Quality-assurance gate registry}
	\label{sec:gates}
	\grainsmith{} evaluates twenty-six construction-time quality-assurance gates
	that re-measure prescribed geometric, structural, and statistical properties
	during generation; their respective evaluation stages are annotated across the
	workflow in Fig.~\ref{fig:pipeline}, while the complete registry and individual
	thresholds are tabulated in \ref{app:gates}. A \emph{hard} gate aborts the run
	with an actionable message, a \emph{warn} gate allows completion but records a
	quality concern in \code{summary.csv}, and a \emph{diagnostic} gate reports a
	quantity without a defensible pass/fail threshold. G20, G23, G25, and G26 are
	diagnostic in this sense and never generate a warning row by themselves. All
	named thresholds are defined in a single constants module with an associated
	written justification; the fixed percentage bands of G8, G9 and G17 are stated
	with the gate itself and in \ref{app:gates}. The gate severity reflects the nature of the check: conditions
	that invalidate the model (such as overlapping atoms or broken symmetry) are
	treated as hard failures, whereas recognised method limitations (such as
	finite-band Hurst offsets or ODF-bounded $\Sigma3$ content) are reported as
	warnings or diagnostics for user interpretation.
	
	Within this registry, gates G22--G26 close the volume-weighted texture and
	domain-warp accounting behind the assignment annealer, whose ODF invariance
	holds in the number-weighted sense only (Sec.~\ref{sec:texture}). G22 re-measures the discrete volume-weighted orientation drift relative to the initial assignment and, when memory permits, separately compares
	a kernel-based discrepancy with a random-assignment reference; G23 reports the residual gap between angle-window and true-CSL $\Sigma3$ area fractions; G24 re-measures the per-grain volume
	deviation introduced by a domain warp after the G11 check on the unwarped
	SDOT base; G25 records configured versus realised texture-component weights;
	and  G26 reports the discrepancy between tessellation-volume and
	atom-count grain weights. Their definitions, thresholds, and operational
	roles are detailed in \ref{app:gates}.
	
	\subsection{Parallel execution model and algorithms}
	\label{sec:parallel}
	\grainsmith{} parallelises the pipeline at the task level, stage by stage,
	subject to a strict determinism requirement: changing the worker count
	must preserve the scientific data byte-for-byte within the fixed
	environment defined in Sec.~\ref{sec:determinism}. This subsection describes the stages
	that are parallelised, their execution mechanisms, and the measures used to
	preserve this contract. Table~\ref{tab:parallel} summarises where
	\code{-{}-jobs} parallelism enters the pipeline of Fig.~\ref{fig:pipeline},
	while Sec.~\ref{sec:performance} reports the measured speedups and identifies
	the serial stage limiting performance for each geometry.

	\begin{table}[!htb]
		\centering
		\caption{Task-level parallelism by pipeline stage.}
		\label{tab:parallel}
		\footnotesize
		\begin{tabular}{@{}p{0.40\linewidth}p{0.52\linewidth}@{}}
			\toprule
			Stage & Parallel mechanism \\
			\midrule
			Config load \& resolve & Serial (master process) \\
			Seeding (RSA + Lloyd) & Serial (master process) \\
			Tessellation & Serial (master process) \\
			Orientation sampling + MDF annealing & Serial (master process) \\
			Atom fill & Process pool over grains (\code{-{}-jobs}); per-grain deterministic RNG substream \\
			Overlap removal + G7 search & Thread-parallel periodic $k$-d-tree ball queries (\code{jobs} scipy threads for \code{jobs}$\,\ge$2; serial pair query at \code{jobs}$\,=$1); canonical lexicographic pair order \\
			Cell-ownership test (curved backends) & Optional numba short-circuit kernel (soft dependency; numpy fallback, \code{GRAINSMITH\_NO\_NUMBA}) \\
			LAMMPS/XYZ atom-section write & Process pool, fixed 250\,000-atom chunks, submission-order drain \\
			Analysis, CSV/summary & Serial (master process) \\
			\bottomrule
		\end{tabular}
	\end{table}
	
	The pipeline integrates established crystallographic, geometric, and statistical
	procedures, whose underlying constructions are described in
	Secs.~\ref{sec:crystal}-\ref{sec:texture}. The implementation adds two
	runtime-specific mechanisms. Post-fill neighbour discovery for overlap removal
	and G7 uses a periodic $k$-d-tree, while the curved-backend cell-ownership test
	uses a short-circuiting nearest-replica argmin implemented as an optional
	numba kernel. The latter is a soft dependency with a verbatim NumPy fallback,
	controlled by \code{GRAINSMITH\_NO\_NUMBA}. When numba is available, the
	kernel is used by default. Both implementations are verified elementwise by
	the test suite, ensuring that the execution path affects runtime only and
	does not alter the generated structure.
	
	Byte-identical scientific data are maintained through three complementary mechanisms.
	First, all stochastic operations are confined to deterministic RNG substreams
	(Sec.~\ref{sec:determinism}). Second, the overlap-removal and G7 pair list is
	sorted into a canonical lexicographic order, making its processing independent
	of discovery order and worker count. Third, the optional numba ownership
	kernel reproduces the NumPy reference implementation's sqrt-space distance
	comparisons and tie-breaking exactly. The repository's end-to-end tests compare the scientific content of
	\code{polycrystal.data} across \code{-{}-jobs} values and between the numba
	and NumPy ownership routes, with provenance timestamps controlled or
	normalised for comparison.
	
	\subsection{Determinism model}
	\label{sec:determinism}
	A single master seed is expanded into named child RNG streams
	(\code{seeding}, \code{orientation}, \code{fields}, \code{occupancy},
	\code{sizes}, \code{mdf}, \code{doping}) using NumPy's
	\code{SeedSequence.\allowbreak spawn()}. The stream list is append-only, so
	adding a new stream does not renumber existing streams. This preserves stream
	assignments across releases, but does not provide cross-version
	reproducibility, since algorithmic changes within a stream may alter its draw
	sequence. Accordingly, byte-level reproducibility is defined for a fixed
	\grainsmith{} version and computational environment, including the
	dependency versions, BLAS build, and CPU architecture.
	
	Direct use of the \code{np.random.*} module functions is forbidden throughout
	the codebase. Each grain uses its own deterministic RNG stream, allowing atom
	filling to be parallelised across worker processes (\code{-{}-jobs}) without
	changing the output. The overlap-removal and G7 neighbour searches follow the
	same principle: their pair lists are processed in canonical lexicographic
	order determined solely by atom coordinates, independent of hash iteration
	order or worker count. Consequently, a fixed configuration and seed reproduce the scientific
	data byte-for-byte across supported worker counts within that environment
	(Sec.~\ref{sec:performance}).
	
	This guarantee distinguishes scientific data from run-dependent metadata.
	Provenance headers and \code{microstructure.json} contain a UTC timestamp;
	setting \code{SOURCE\_DATE\_EPOCH} fixes that timestamp to a specified
	instant. Execution records, including worker-count metadata, timing or
	memory measurements, and \code{run.log}, are outside the scientific-data
	identity guarantee. Hashes in \code{MANIFEST.txt} reflect the actual file
	contents, so files containing different metadata can have different
	digests despite identical scientific data. Timestamp control alone
	therefore does not establish byte identity of the entire output directory.
	
	Cross-platform execution is expected to be numerically equivalent but is not
	guaranteed to be bit-identical. Differences in last-bit arithmetic can
	occasionally cross a geometric decision threshold, for example when an atom
	lies on the opposite side of a cell face or an overlap pair lies at the
	cutoff, producing a discrete geometric difference rather than a
	rounding-level variation.
	
	\subsection{Interfaces and outputs}
	\label{sec:interfaces}
	The package is driven through a command-line interface using a YAML
	configuration validated by the engine's resolver. A typical session uses two
	commands: \code{grainsmith validate config.yaml} checks the configuration
	without building anything (gate G1, schema and cross-field rules, and a G2
	crystal dry build), while \code{grainsmith generate config.yaml} executes the
	full pipeline and writes the outputs described below. The
	\code{grainsmith info} subcommand provides space-group and Wyckoff reference
	information for all \sgnum{} groups. Ready-to-run, commented configurations
	for each generation method are provided under \code{examples/}, with their
	repository paths catalogued in \ref{app:examples}; configuration, geometry
	and physics conventions, and the gate registry are documented under
	\code{docs/}.
	
	Each run produces LAMMPS data and extended XYZ files, together with
	per-grain and per-boundary CSV files (\code{grains.csv},
	\code{boundaries.csv}, and \code{vertices.csv} for flat and power
	geometries), a disorientation-angle
	histogram (\code{mdf.csv}) against the Haar-random reference,
	MTEX-ready orientations (\code{odf\_mtex.txt}), goodness-of-fit statistics
	(\code{statistics.csv}), and a machine-readable record
	(\code{microstructure.json}) containing configuration, provenance,
	environment, and analysis information. Additional outputs include
	\code{summary.csv}, which records QA results and execution metadata,
	a gnuplot bundle, and a SHA-256 \code{MANIFEST.txt}. The \code{mdf.csv}
	filename is historical: it contains the boundary-area-weighted
	disorientation-angle marginal, not a full five-parameter
	boundary-character distribution. Multiphase runs replace the single-phase
	orientation/MDF files with per-phase ones and add interphase-boundary columns,
	whereas doped runs add per-grain \code{doping.csv}. Opt-in curvature
	analysis adds curvature columns to \code{boundaries.csv} and local
	samples to \code{gb\_curvature.csv} (Sec.~\ref{sec:analysis}).
	
	The MD reproducibility and FAIR literature~\cite{mdreproducibility2023,mosdef2025}
	identifies three recurring requirements: machine-readable provenance, a
	complete configuration record, and deterministic regeneration. Within the
	atomistic-generation tools surveyed in Sec.~\ref{sec:motivation}, these
	requirements are not commonly exposed as a single explicit generation
	contract. \grainsmith{} implements these requirements through deterministic
	seed-to-stream assignment, machine-readable provenance in
	\code{microstructure.json}, file-integrity records in
	\code{MANIFEST.txt}, and a complete resolved configuration in
	\cfg{resolved_config.yaml}. Scientific-data reproducibility follows
	the environment and metadata conditions in Sec.~\ref{sec:determinism}.
	The \code{statistics.csv} additionally reports goodness-of-fit values together
	with their documented estimator biases, enabling users to interpret the
	reported statistics in light of their known limitations.
	
	% =====================================================================
	\section{Microstructure generation}
	\label{sec:theory}
	
	This section follows the generation pipeline from crystal construction through
	tessellation, texture, multiphase partitioning, dopant insertion, analysis, and
	quality-assurance gates. For each method, we describe its construction,
	verification, and documented limits. Throughout, \emph{exact} and
	\emph{exactly} denote constructional or combinatorial identities, such as
	half-space intersections, integer lattice multiples, and permutations of fixed
	orientation sets, and never floating-point equality. Quantities subject to
	finite-precision arithmetic are therefore reported with the tolerance of the
	corresponding gate. Configured targets are treated as specifications rather
	than guarantees; their realised values are established by construction-time
	re-measurement, with relevant deviations reported. Ready-to-run, commented
	examples for each method are catalogued in \ref{app:examples}.
	
	\subsection{Space-group-aware crystal generation}
	\label{sec:crystal}
	A crystal can be supplied either through a space-group number, lattice
	parameters, and Wyckoff sites, with optional partial occupancy, or
	through a CIF file. For explicit space-group/Wyckoff input,
	\grainsmith{} expands the specified orbits into a unit cell and verifies
	its symmetry using spglib~\cite{togo2018spglib}; the detected
	International Tables for Crystallography (ITA) space-group number must
	match the requested number. For CIF input, spglib first detects the
	symmetry of the imported structure at the configured \code{symprec}.
	The symmetry-inequivalent Wyckoff sites are then expanded from its
	standardised cell, and the reconstructed cell is checked against this
	detected group. G2 is a hard round-trip check on both routes. A mismatch
	between the CIF header's declared group and the initially detected group
	produces a separate warning, rather than a G2 failure. Accordingly,
	\code{requested\_sg} in the CIF-path report denotes the detected reference
	group, not the declaration in the file header. This verification concerns
	the reference unit cell and does not establish the global symmetry of
	the assembled polycrystal or its interfaces.
	
	An exhaustive per-group sweep using one generic Wyckoff decoration was
	performed through the production code path and independently re-detected with
	spglib at $10^{-4}$ symprec. All \sgnum{} groups passed: 177 with a
	single-species decoration and 53 with a two-species decoration introduced to
	remove accidental supergroup symmetry. This sweep verifies one representative
	decoration per group; it does not exhaustively cover all possible Wyckoff
	combinations, special positions, free coordinates, partial occupancies,
	settings, or chemical decorations. The corresponding per-group records are
	provided as supplementary data, while the shipped examples cover both manual
	Wyckoff specifications and direct CIF input, including a case with
	inconsistent declared and detected symmetry (\ref{app:examples}).
	
	\subsection{Tessellation}
	\label{sec:tessellation}
	
\begin{figure*}[!t]
		\centering
		\begin{subfigure}[t]{0.25\linewidth}
			\IfFileExists{figures/fig2_a_geometries.pdf}%
			{\includegraphics[width=\linewidth]{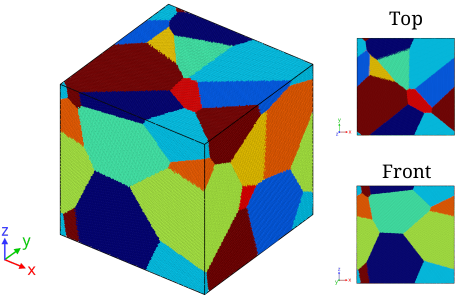}}%
			{\fbox{\parbox[c][0.62\linewidth][c]{0.94\linewidth}{\centering
						OVITO composite\\
						\code{fig\_2/a/polycrystal.extxyz}}}}
			\caption{}\label{fig:geometries-a}
		\end{subfigure}\hspace{0.8cm}
		\begin{subfigure}[t]{0.25\linewidth}
			\IfFileExists{figures/fig2_b_geometries.pdf}%
			{\includegraphics[width=\linewidth]{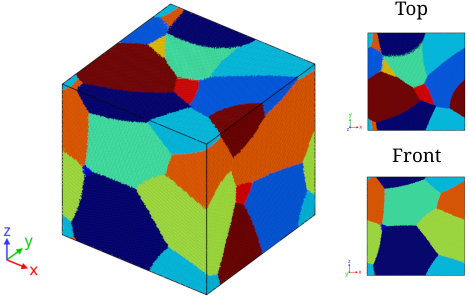}}%
			{\fbox{\parbox[c][0.62\linewidth][c]{0.94\linewidth}{\centering
						OVITO composite\\
						\code{fig\_2/b/polycrystal.extxyz}}}}
			\caption{}\label{fig:geometries-b}
		\end{subfigure}\\[2pt]
		\begin{subfigure}[t]{0.25\linewidth}
			\IfFileExists{figures/fig2_c_geometries.pdf}%
			{\includegraphics[width=\linewidth]{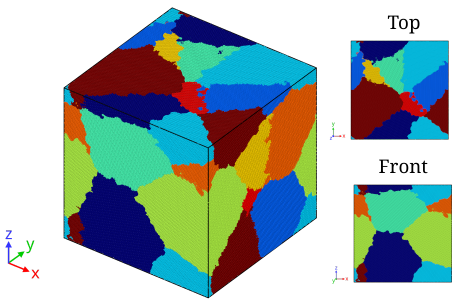}}%
			{\fbox{\parbox[c][0.62\linewidth][c]{0.94\linewidth}{\centering
						OVITO composite\\
						\code{fig\_2/c/polycrystal.extxyz}}}}
			\caption{}\label{fig:geometries-c}
		\end{subfigure}\hspace{0.8cm}
		\begin{subfigure}[t]{0.25\linewidth}
			\IfFileExists{figures/fig2_d_geometries.pdf}%
			{\includegraphics[width=\linewidth]{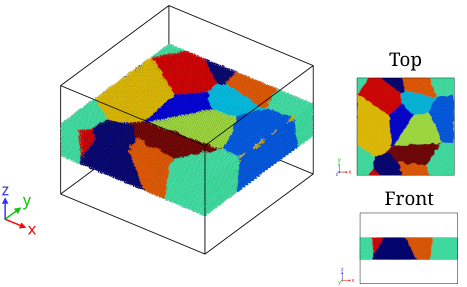}}%
			{\fbox{\parbox[c][0.62\linewidth][c]{0.94\linewidth}{\centering
						OVITO composite\\
						\code{fig\_2/d/polycrystal.extxyz}}}}
			\caption{}\label{fig:geometries-d}
		\end{subfigure}
		\caption{Grain-boundary geometry families, rendered in
			OVITO~\cite{stukowski2010ovito} from the extended-XYZ outputs (atoms coloured
			by grain). (a)~Flat Voronoi boundaries: FCC Cu, 10 grains, $(360~\text{\AA})^3$
			periodic box. (b)~Curved boundaries, anisotropic (ellipsoidal-metric)
			backend, on the same Cu system. (c)~Self-affine rough boundaries,
			\code{perturbed\_distance} spectrum with Hurst exponent $\mathcal{H}=0.8$.
			(d)~The same self-affine construction on a CIF-sourced anatase TiO$_2$
			($I4_1/amd$, detected by spglib from a P1 export) thin film,
			a material slab of $360\times360\times80$~\AA{} with free $z$ surfaces; \code{box.vacuum}$=180$~\AA{} is the total vacuum added on that axis, i.e.\ $90$~\AA{} beyond each surface, so the exported box is $360\times360\times260$~\AA{}. The atomic structure shown in each panel can be regenerated from its shipped YAML configuration under the reproducibility conditions of Sec.~\ref{sec:determinism}.}
		\label{fig:geometries}
	\end{figure*}

	The default tessellation is periodic Voronoi. Planar faces are constructed
	with Qhull from periodic translation replicas; along free axes, mirrored seed
	images clip cells at the walls through a $d^2$ half-space inequality. The
	\code{owns()} home-cell rule partitions the periodic domain exactly once, so
	filling each compact home cell emits every torus point once, including for
	lattices incommensurate with the simulation box. No wrap-and-deduplicate step
	is therefore required, avoiding the double counting that can occur when
	wrapped membership is combined with rounding-based deduplication.
	Fig.~\ref{fig:geometries-a} shows the resulting flat-faced polycrystal for a
	10-grain FCC Cu system; the \code{basics} group provides the corresponding
	minimal configuration (\ref{app:examples}).
	
	To prescribe grain volumes, \grainsmith{} uses volume-targeted Laguerre
	(power) diagrams. The power weights $w$ are fitted by periodic SDOT through
	damped Newton iterations on the Kantorovich dual~\cite{kitagawa2019newton}, following the polycrystalline framework established by Bourne et al ~\cite{bourne2020laguerre,bourne2023sdot}. For cell
	volumes $V(w)$ and targets $V^\ast$, the residual is
	$F(w)=V(w)-V^\ast$, with Jacobian
	\begin{equation}
		J_{ij}=\frac{\partial V_i}{\partial w_j}
		=-\frac{A_{ij}}{2 d_{ij}}\quad(i\neq j),
		\qquad
		J_{ii}=\sum_{j\neq i}\frac{A_{ij}}{2 d_{ij}},
		\label{eq:sdot}
	\end{equation}
	where $A_{ij}$ is the shared radical-face area obtained directly from the
	Qhull/scipy half-space-intersection polyhedron through its
	\texttt{ridge}/face areas, and $d_{ij}$ is the seed distance. The Jacobian is
	therefore fully analytic and has the structure of a weighted graph Laplacian
	on the grain-adjacency graph, with conductance $A_{ij}/(2d_{ij})$. It is
	symmetric; its vanishing column sums express conservation of the fixed box
	volume ($\sum_i V_i=V_{\mathrm{box}}$), while the corresponding vanishing row
	sums express the invariance of the diagram under a uniform weight shift. The
	gauge is therefore fixed by $w_0=0$.
	
	Damping backtracks the Newton step to prevent cell collapse. In all reported
	configurations, the iteration reaches the specified volume tolerance within
	3--8 Newton steps
	(Secs.~\ref{sec:validation-structural} and~\ref{sec:performance}). For arbitrary seed arrangements or target spreads, convergence cannot be guaranteed. The algorithm terminates with an error if step sizes exceed the backtracking limit, while reaching the iteration budget simply returns the residual, leading to a hard gate G11 failure upon re-measuring per-grain volume discrepancy. An optional
	centroidal loop~\cite{kuhn2020centroidal} moves seeds to power-cell centroids
	for equiaxed grains with prescribed volumes. The analytic Jacobian
	(Eq.\eqref{eq:sdot}) is independently verified against finite differences to a
	relative error of $\sim10^{-9}$, while its zero-row-sum structure and the
	quadratic-convergence regime near the solution are recovered in the test
	suite.
	
	The isotropic case ($M_i=\mathrm{Id}$, where $M_i$ is grain $i$'s anisotropy
	metric and \texttt{M} in the implementation) is the corresponding special case
	of the anisotropic power-diagram framework of Buze et al.~\cite{buze2024pyapd}.
	\grainsmith{} performs the fit on explicitly reconstructed polyhedral cells
	using Qhull half-space intersections, yielding the analytic Jacobian in
	(Eq.\eqref{eq:sdot}) and a tight volume tolerance of \texttt{vol\_tol} $\sim 10^{-3}$ (0.1\% per cell) without a voxel discretisation.
	The \texttt{grain\_geometry} group provides log-normal, equal-volume, explicit
	volume-list, and centroidal configurations (\ref{app:examples}).

	\subsection{Curved and self-affine grain boundaries}
	\label{sec:curved}
	
	Grain boundaries need not remain flat. A periodic Gaussian random field warps
	the diagram (the \code{warp} method), subject to bijectivity guards (gate G6:
	displacement-gradient norm $<0.5$, with amplitude tied to
	$\code{min\_seed\_distance}/4$). The \code{additive\_weights}
	(Johnson--Mehl/Apollonius) and \code{anisotropic} (ellipsoidal-metric)
	backends provide alternatives. The warp curves the faces while preserving
	Voronoi topology, i.e.\ it does not create new neighbours. The \code{additive\_weights} and \code{anisotropic} backends instead reshape the cells through a different local metric and can change neighbours. Genuinely non-Voronoi shapes come from the two methods that replace the assignment rule itself: \code{perturbed\_distance} for self-affine boundaries, and \code{voxel\_import} (the \code{voxel\_import} group) for externally supplied label fields.
	
	Unlike the anisotropic power diagram of~\cite{buze2024pyapd}, the
	\code{anisotropic} backend is a Voronoi-class metric with $w_i=0$:
	\grainsmith{} does not couple the anisotropy matrices to the SDOT
	volume fit. Prescribed grain volumes (flat power faces) and
	anisotropic or curved shape are therefore separate backends, rather
	than components of a single volume-targeted APD. They can nevertheless be
	composed: requesting a \code{size\_distribution} together with a
	\code{warp} first builds the SDOT power diagram and then warps it. Volume
	targeting is consequently enforced on the unwarped power base:
	gate G11 re-measures the base cells, and its message states this explicitly,
	because the warp deliberately displaces the boundaries and therefore
	perturbs the realised per-grain volumes, bounded by the G6
	displacement-gradient guard. So that the composed case is not left
	unmeasured, gate G24 re-measures the per-grain volumes on the
	warped geometry and warns when the relative deviation from the target
	exceeds \code{WARP\_VOLUME\_G24\_TOL}$=0.10$. It runs only when the warp
	geometry differs from the SDOT base that G11 has already checked, so the volume deviation introduced by the warp is measured and reported on the structure that is actually exported.
	
	The two tolerances are deliberately of different kinds: \code{vol\_tol} is a fit precision on the unwarped base, whereas $0.10$ is a loose sanity bound on a displacement that exceeds that precision by orders of magnitude. Calibrated at warp amplitude $0.6$ in a $(60~\text{\AA})^3$ box with a $10~\text{\AA}$ correlation length, the post-warp maximum relative volume error was $0.021$ at four grains and $0.041$ at eight, against $1.5\times10^{-4}$ and $6.4\times10^{-5}$ on the base geometry that G11 measures. For equiaxed grains with tight realised
	volumes, the centroidal loop should therefore be used instead of a warp.
	Fig.~\ref{fig:geometries}(b) shows the resulting gently bowed, elongated
	grains of the \code{anisotropic} backend on a 10-grain FCC Cu system. The
	\code{grain\_geometry} group ships one configuration for each curved backend,
	including the single legal curved-plus-prescribed-volume combination
	(\ref{app:examples}).
	
	For self-affine boundaries, the \code{perturbed\_distance} method does not displace a finished diagram at all: it perturbs the assignment rule itself, so that the grain label at a point $x$ is $\arg\min_i\left[d_{\mathrm{pbc}}(x,s_i)-A_\eta\,\eta_i(x)\right]$. Every grain $i$ carries its own independent scalar random field $\eta_i$, drawn from a band-limited power-law amplitude spectrum
	$\sqrt{S(k)}\propto k^{-(3+2\mathcal{H})/2}$ over
	$k\in[2\pi/l_{\max},2\pi/l_{\min}]$. Each such field has power $P(k)\propto k^{-(3+2\mathcal{H})}$, while its planar restriction has surface
	PSD $C(q)\propto q^{-2-2\mathcal{H}}$. For an ideal self-affine height surface, this spectral convention
	corresponds to $D_s=3-\mathcal{H}$~\cite{jacobs2017psd}.
	This relation does not by itself establish the fractal dimension of
	the grain boundaries produced by the assignment rule. We use $\mathcal{H}$ exclusively for the Hurst
	exponent and reserve $H$ for the local mean curvature shown in
	Fig.~\ref{fig:curvature}, matching the \code{H\_invA} column written by the
	analysis stage. The field amplitude is likewise written $A_\eta$
	throughout, leaving $A$ for the unrotated lattice matrix of
	Sec.~\ref{sec:multiphase}.
	
	Self-affine GB morphology at the nanoscale is experimentally established
	\cite{braun2018fractal}, while the fractal-surface construction used for MD
	follows~\cite{eder2017abrasion}. Gate G13 back-estimates $\mathcal{H}$ from
	the generated field using a radial-PSD log-log fit. Because a finite or under-resolved scale-free band can bias this fit, G13 is a \emph{warn} gate, and its tolerance \code{HURST\_G13\_TOL}$=0.15$ sits above the documented bias of the estimator itself. Shell discreteness at the band edges and the per-mode $\chi^2$ scatter of a single realisation displace the fitted slope by up to $\sim0.1$ for a one-decade band.
	For a well-sampled band, an independent radially averaged PSD reproduces the
	G13 estimate to all printed digits and recovers $\mathcal{H}$ to $<0.005$ for
	$\mathcal{H}\in\{0.5,0.7,0.9\}$.
	
	G13 and G20 assess different objects. G13 back-estimates the Hurst
	exponent of the synthesised random field. G20 instead fits a
	box-counting roughness index $D_b$ to grain perimeters in two-dimensional
	sections of the resulting geometry, following the section-based
	approach of Braun et al.~\cite{braun2020fractal}. It samples up to three
	in-box sections of each of the five largest grains by voxel count,
	after connectivity repair; G19 reports the fraction of reassigned
	voxels. The fit uses $\log N(\epsilon)$ versus $\log(1/\epsilon)$,
	with box sizes restricted to the synthesis band
	$\epsilon\in[l_{\min},l_{\max}]$ at the section's pixel scale.
	Thus, $D_b$ describes section perimeters rather than three-dimensional
	boundary surfaces and is not interchangeable with $D_s$.
	
	The finite-resolution estimator requires calibration. At the documented
	200-pixel section resolution, straight-edge tests give
	$D_b\approx1.027$, while rasterised disk perimeters give approximately
	$1.07$--$1.15$, depending on radius and perimeter extraction.
	An unperturbed ($A_\eta=0$) reference gives approximately $1.07$.
	Consequently, a value above unity alone does not establish self-affinity.
	
	The documented bandwidth calibration also distinguishes statistical
	significance from effect magnitude. At approximately 3.2 synthesis-band
	octaves, associated with $\sim10$~nm grains in that calibration, the
	Hurst-dependent response has
	$\eta^2\approx0.29$--$0.34$ and $p\approx5\times10^{-6}$.
	An effect size $\eta^2>0.8$ is reached near 5.5 octaves
	($\sim50$~nm grains under those settings). The narrower
	$\sim1.7$-octave validation configuration gives
	$\eta^2\approx0.06$ and $p\approx0.14$, providing insufficient evidence
	to resolve a Hurst-dependent response in that test.
	These are calibration-specific observations, not universal grain-size
	thresholds or guarantees of a converged fractal dimension.
	
	For the Cu structure in Fig.~\ref{fig:geometries}(c),
	G20 reports $D_b=1.114$, while G13 recovers $\mathcal{H}=0.800$.
	The former lies within the individual-grain range
	$1.10\le D_b\le1.24$ reported for annealed
	Pd$_{90}$Au$_{10}$, below its 57-grain mean of
	$1.174\pm0.004$~\cite{braun2020fractal}.
	This numerical overlap is contextual rather than a validation:
	the materials, resolved scales, and measurement conditions differ,
	and the present estimator has a documented rasterisation bias.
	G20 therefore remains a threshold-free diagnostic of resolved
	section-perimeter roughness; it reports no estimate when no sampled
	section supports a usable fit. The specific GB area $S_V$, defined as
	boundary area per unit volume, provides a complementary morphological
	measure, subject to the voxel-area bias described in
	Sec.~\ref{sec:analysis}.

	Fig.~\ref{fig:geometries}(c) shows the resulting rough boundaries
	(\code{perturbed\_distance}, $\mathcal{H}=0.8$) on the same Cu system, while
	Fig.~\ref{fig:geometries}(d) shows the same construction on a CIF-sourced
	anatase TiO$_2$ thin film with free surfaces. The
	\code{self\_affine\_gb} group ships the Pd--Au reference case and the
	thin-film case of panel~(d) (\ref{app:examples}).
	
	\subsection{Texture (ODF) and boundary disorientation-angle statistics}
	\label{sec:texture}
	
	\grainsmith{} uses active, scalar-first unit quaternions
	$q=(q_0,q_1,q_2,q_3)$ encoding
	$v_{\mathrm{lab}}=R(q)\,v_{\mathrm{crystal}}$. The standard Bunge orientation
	matrix maps sample $\rightarrow$ crystal,
	$g=R(q)^{\!\top}=[\,\mathrm{ZXZ}(\varphi_1,\Phi,\varphi_2)\,]^{\top}$, so the
	exported Bunge angles are extracted from $R(q)$. This matches the
	MTEX/EBSD convention~\cite{engler2009texture} and is pinned against the
	textbook Bunge matrix and the ideal Brass component
	$\{011\}\langle 2\bar{1}1\rangle
	\rightarrow(35.26^\circ,45^\circ,0^\circ)$.
	
	The ODF is constructed from weighted components: an Euler centre with
	isotropic spread (an exact Haar-corrected small-angle law via rejection
	sampling, using uniform SO(3) sampling~\cite{shoemake1992rotations}), a
	fibre component, or a Haar-uniform random fraction.
	Because the ODF is a volume-weighted density, each component weight is by
	default realised as a volume fraction (\code{component\_weight\_basis:
		volume}): grains are partitioned deterministically so that the realised volume fractions approach the configured weights as closely as the discrete grain volumes allow. The alternative \code{count} basis
	draws each grain's component from a categorical distribution proportional to
	the weights, realising them as grain-count fractions instead; the two
	bases coincide only when all grains have equal volume. Gate G25 records the
	configured weight beside both realised fractions for every component.
	The boundary network is
	then shaped through \code{mdf\_target}: simulated annealing
	permutes which grain receives which orientation, so the multiset of sampled orientations, and with it the
	number-weighted orientation distribution is conserved exactly. This
	is the conserved-texture variant of the Monte-Carlo swap algorithm of
	Miodownik et al.~\cite{miodownik1999bmd}, whose lattice domains are all
	the same size. Grains are not: a swap exchanges the volumes carried by two
	orientations, so the volume-weighted ODF is in general not invariant.
	Saylor et al.~\cite{saylor2004statistically} reach the same conclusion
	from the opposite direction, observing that an orientation distribution left
	out of the swap objective is unlikely to survive the optimisation.
	\grainsmith{} therefore measures this drift rather than asserting its absence.
	The optional \code{odf\_drift\_max} bounds the change in the discrete
	volume-weighted orientation distribution relative to the sampler's
	initial assignment. Let $\pi_0$ and $\pi$ denote the initial and current
	assignments, and let $c$ index distinct orientation classes after
	grouping duplicate and crystal-symmetry-equivalent orientations. The
	controlled quantity is
	\begin{equation}
		D_{\mathrm{TV}}(\pi,\pi_0)
		=
		\frac{1}{2}\sum_c
		\left|w_c(\pi)-w_c(\pi_0)\right|,
		\qquad
		w_c(\pi)
		=
		\frac{\sum_{i:\,\pi(i)\in c}V_i}{\sum_i V_i},
		\label{eq:odf-drift}
	\end{equation}
	where $V_i$ is the tessellation volume of grain $i$.
	Every accepted swap must satisfy the configured cap; when the cap is
	unset, drift is measured but not constrained. G22 re-measures this
	quantity in the analysis stage independently of the annealer's
	incremental bookkeeping. The constraint bounds changes introduced by
	reassignment, not the initial sampling error relative to a prescribed
	continuous ODF.
	
	When memory permits, G22 also evaluates a maximum mean discrepancy
	(MMD) using a symmetry-averaged de la Vall\'ee Poussin kernel. Its
	half-width is configured through \code{odf\_kernel\_halfwidth\_deg}
	(default $10^\circ$), with the effective width reported after
	integer-degree selection. The final volume-weighted distribution is
	compared with the count-weighted reference, and the resulting MMD is
	assessed against the central 95\% interval from random assignments of
	the same grain volumes to the fixed orientation set
	(\code{odf\_null\_samples}, default 256).
	A separate MMD relative to the initial assignment is reported
	descriptively. These diagnostics do not enter the annealing constraint:
	normalised kernel smoothing cannot increase total-variation drift, but
	the same numerical cap does not apply to MMD. The finite-bandwidth MMD
	is a pseudometric, so a zero value alone does not establish equality of
	the physical ODFs.
	
	Gate G26 complements the orientation-drift measurement by comparing
	tessellation-volume and atom-count weights on the same grain indices:
	\begin{equation}
		\Delta_{\mathrm{VA}}
		=
		\frac{1}{2}\sum_i
		\left|
		\frac{V_i}{\sum_j V_j}
		-
		\frac{n_i}{\sum_j n_j}
		\right|,
		\label{eq:weighting-gap}
	\end{equation}
	where $n_i$ is the final atom count of grain $i$. It also reports the
	maximum absolute per-grain weight difference and the mean atom count
	per grain. Multiphase runs are assessed separately within each phase
	to avoid conflating differences in atomic number density with
	within-phase discretisation effects. Atom-count weighting is a useful
	proxy for volume weighting when atomic volumes are equivalent, but
	does not generally replace the physical volume-weighted ODF.
	
	Unlike Eq.~\eqref{eq:odf-drift}, Eq.~\eqref{eq:weighting-gap} compares
	weights on individual grain indices. Grouping grains with identical
	or symmetry-equivalent orientations can reduce this discrepancy;
	$\Delta_{\mathrm{VA}}$ is therefore an upper bound on the corresponding
	difference between the grouped orientation distributions. It is
	reported as a diagnostic without a pass/fail threshold.
	
	In the documented fixed-grain-count size sweep, the discrepancy falls
	from $0.027$ to $0.0086$ over 275--9191 atoms per grain, with a fitted
	power-law exponent of $-0.33$. This trend is consistent with a
	boundary-shell contribution whose relative magnitude decreases with
	grain size. It provides an empirical discretisation scale for the
	tested configurations, rather than a universal lower bound on ODF
	accuracy. Tightening \code{odf\_drift\_max} controls reassignment-induced
	drift in the tessellation-volume weighting; it does not directly
	constrain the separate volume-versus-atom-count discrepancy measured
	by G26.
	
	The quantity this procedure controls is the boundary-area-weighted
	disorientation-angle distribution: a one-dimensional marginal of the
	full misorientation distribution, which requires three parameters to specify a misorientation and five to specify a boundary~\cite{saylor2004statistically}. The objective prescribes neither the misorientation axis nor the boundary-plane character, and should therefore not be read as control of the full five-parameter grain-boundary character distribution. The angle distributions of Ref.~\cite{miodownik1999bmd} are marginals in the same sense, and the distinction is not cosmetic where $\Sigma3$ is concerned.
	
	Three target forms are available. \code{haar\_random} (deprecated alias \code{mackenzie}) is the random-pair
	reference of the crystal point group, which in the cubic case is the Mackenzie
	law~\cite{mackenzie1958misorientation}. \code{sigma3\_angle\_enriched} (deprecated alias \code{csl\_enriched})
	starts from that base and moves a target $\Sigma3$ mixture fraction into
	the angular Brandon window $|\theta-60^\circ|\le15^\circ/\sqrt3$ alone~\cite{brandon1966csl}, where \code{sigma3\_fraction} (written $f_{\Sigma3}$ below and in
	Fig.~\ref{fig:validation}) is the mixture coefficient of that construction rather than the final area fraction inside the window, because the Haar-random component already places mass there. The third form is an explicit target histogram. Gate G12 reports the final $\chi^2$ distance, and gate
	G23 places the area fraction inside that angular window beside the area
	fraction of true $\Sigma3$ boundaries, which additionally satisfy the
	$\langle111\rangle$ axis condition. Membership of the angular window is
	necessary but not sufficient for $\Sigma3$ character, so the true CSL $\Sigma3$ area fraction is necessarily no larger than the corresponding angular-window fraction.
	
	The Haar-random reference is not evaluated in closed form,
	and no single closed-form expression analogous to the cubic Mackenzie law is used here for non-cubic point groups. Instead, it is
	generated once per point group as a seeded $2\times10^{5}$-pair Haar-random
	Monte-Carlo curve and is therefore a deterministic function of the point
	group alone, rather than of the run seed. The name Mackenzie is
	reserved in the code and in this text for the cubic case it describes. For $m\bar{3}m$, the reference reproduces the characteristic
	features of the cubic Mackenzie law, including region masses
	of approximately 0.598 below $45^\circ$ and 0.394 in
	$[45^\circ,60^\circ)$, and a median near $42.3^\circ$.
	The discrepancy between the realised and target histograms
	is measured by the symmetric $\chi^2$ distance,
	\[
	\chi^2 = \sum_b \frac{(p_b-t_b)^2}{p_b+t_b} \in [0,2],
	\]
	where $p_b$ and $t_b$ are normalised bin masses and bins empty
	in both distributions contribute zero.
	Gate G12 uses a configurable warning threshold
	(\code{chi2\_max}, default 0.5), which is an operational
	tolerance rather than a calibrated statistical significance
	level. Unequal boundary-area weights and shared grain
	orientations preclude interpreting the bin-to-pair count
	ratio alone as a universal sampling-noise floor.
	
	The scope of disorientation-angle control is limited by the annealing mechanism itself.
	Because it only permutes a fixed set of orientations over the fixed
	Voronoi neighbour graph, it shapes the angle distribution toward the target rather
	than matching it exactly. That a fixed orientation set brackets what
	permutation can reach is a property of the method, not of this implementation:
	for a random texture on a lattice, Miodownik et al.\ obtain mean
	disorientation angles of $18.2^\circ$ and $47.9^\circ$ for the two extreme
	conserved-texture arrangements, against $40.7^\circ$ for the random
	arrangement~\cite{miodownik1999bmd}. The achievable $\Sigma 3$ area fraction is bounded
	both by the ODF's intrinsic twin content and by the topology of the boundary
	network. Empirically (Sec.~\ref{sec:validation}), across a $0.10$--$0.50$ target sweep, the achieved Brandon-window mass falls increasingly short of the target, while the CSL $\Sigma 3$ area fraction
	remains below $0.06$. The achievable enrichment depends on the supplied orientation set and
	the grain-adjacency graph, while the value reached in a finite run also
	depends on the optimisation schedule. Users can supply twin-related
	ODF components, as demonstrated by the shipped twin-pair example.
	The assignment annealer itself neither creates new orientations nor
	changes the grain-adjacency graph. Thus, \code{mdf\_target} shapes the disorientation-angle
	distribution subject to the fixed orientation set and
	grain-adjacency graph; G12 reports the remaining target
	discrepancy as a warning diagnostic. The \code{texture} group ships the twin-pair ODF with a \code{sigma3\_angle\_enriched} target
	and a fibre-textured thin film (\ref{app:examples}).

	\begin{figure*}[!b]
		\centering
		\begin{subfigure}[t]{0.25\linewidth}
			\IfFileExists{figures/fig3a_enrich20.pdf}%
			{\includegraphics[width=\linewidth]{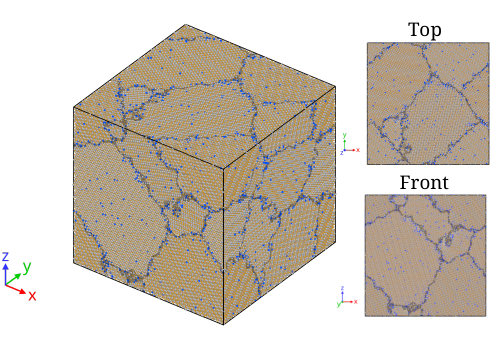}}%
			{\fbox{\parbox[c][0.62\linewidth][c]{0.94\linewidth}{\centering OVITO:
						$E=20$ GB-enriched Li\\ \code{fig\_3/20\_enrich\_}\allowbreak\code{tio2\_li\_dopant}}}}
			\caption{}\label{fig:doping-a}
		\end{subfigure}\hspace{3.1cm}
		\begin{subfigure}[t]{0.25\linewidth}
			\IfFileExists{figures/fig3b_enrich100.pdf}%
			{\includegraphics[width=\linewidth]{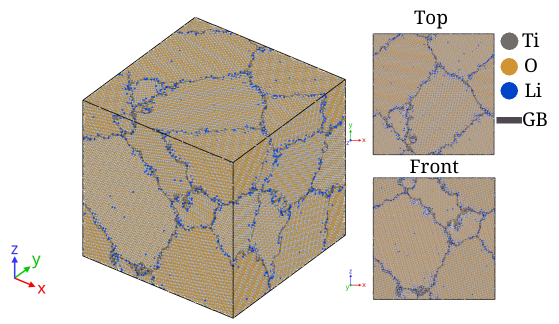}}%
			{\fbox{\parbox[c][0.62\linewidth][c]{0.94\linewidth}{\centering OVITO:
						$E=100$ GB-enriched Li\\ \code{fig\_3/100\_enrich\_}\allowbreak\code{tio2\_li\_dopant}}}}
			\caption{}\label{fig:doping-b}
		\end{subfigure}\\[2pt]
		\begin{subfigure}[t]{0.48\linewidth}
			\includegraphics[width=\linewidth]{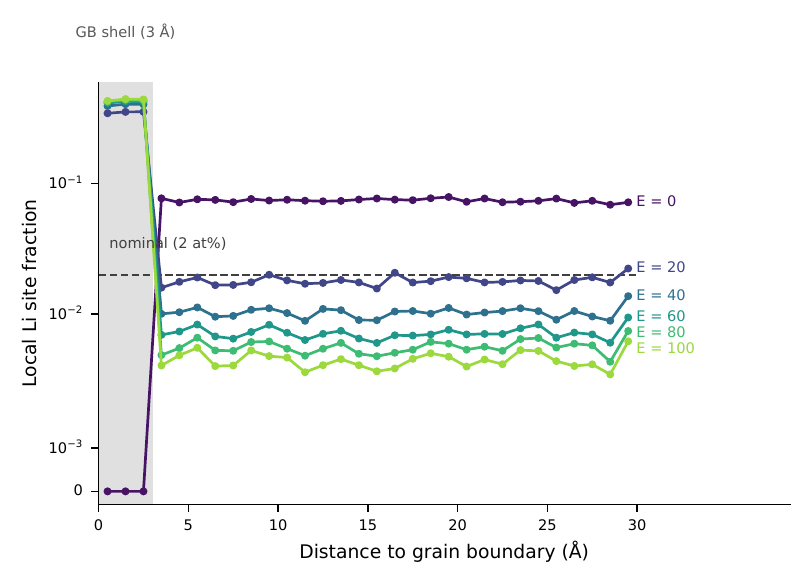}
			\caption{}\label{fig:doping-c}
		\end{subfigure}\hspace{0.1cm}
		\begin{subfigure}[t]{0.30\linewidth}
			\IfFileExists{figures/fig3d_multiphase_tio2.pdf}%
			{\includegraphics[width=\linewidth]{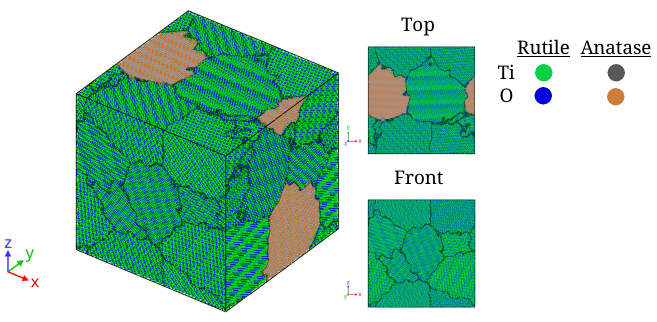}}%
			{\fbox{\parbox[c][0.62\linewidth][c]{0.94\linewidth}{\centering OVITO:
						anatase/rutile two-phase\\ \code{fig\_3/multiphase\_tio2\_}\allowbreak\code{anatase\_rutile}}}}
			\caption{}\label{fig:doping-d}
		\end{subfigure}
		\caption{Grain-boundary doping and multiphase construction on TiO$_2$,
			rendered in OVITO from the extended-XYZ outputs.
			(a)~Anatase polycrystal ($(260~\text{\AA})^3$, 6 grains, self-affine
			boundaries, $\mathcal{H}=0.8$) with 2~at\% Li interstitials in a GB shell of
			3~\AA{} half-thickness at $E=20$: $22\,471$ of $30\,069$ Li occupy the shell, with achieved
			$E=19.9$. (b)~The same host at $E=100$: $28\,052$ of $29\,959$ Li occupy the
			shell. (c)~Achieved local Li site fraction, defined as the placed-dopant count
			divided by the eligible-candidate count in each GB-distance bin, for
			$E\in\{10^{-5},20,40,60,80,100\}$. The shaded band marks
			$d\le3$~\AA{}, corresponding to the configured shell half-thickness.
			The dashed 2~at\% line denotes the target dopant fraction among all
			atoms and therefore uses a different denominator from the plotted
			site fraction. For the reported $E=20$ realisation, the overall
			candidate-site occupancy is $30\,069/498\,951\approx0.060$.
			The $E=20$--$100$ series uses a common 6-grain host; the
			$E=10^{-5}$ shell-depletion example uses a separate 10-grain host.
			The no-preference condition $E=1$ is not shown.
			(d)~Two-phase anatase ($I4_1/amd$) / rutile ($P4_2/mnm$) polycrystal with 50/50
			target volume fractions (10 grains, $(260~\text{\AA})^3$,
			$1\,570\,159$ atoms), using the same curved-boundary construction.}
		\label{fig:doping}
	\end{figure*}	
	
	\subsection{Dopant insertion}
	\label{sec:doping}
	
	An optional \code{doping:} stage, available for single-phase configurations,
	inserts dopants deterministically after overlap removal using a dedicated
	per-grain \code{doping} RNG stream family. The doped atomic configurations therefore remain byte-identical across
	supported \code{-{}-jobs} values under the conditions of Sec.~\ref{sec:determinism}, while undoped configurations are unaffected. Substitutional dopants replace host atoms in place, whereas interstitial dopants occupy crystallographic interstices of each grain's own
	lattice and orientation using named site presets (octahedral\/tetrahedral,
	fcc\/bcc\/hcp) or explicit fractional coordinates. Candidate sites are rejected
	when they violate the per-dopant minimum-distance cutoff, with PBC taken into
	account.
	
	GB-targeted placement uses a two-level shell/bulk Bernoulli model.
	The configured dopant concentration is a target atom fraction of the
	final structure, whereas enrichment is defined using candidate-site
	occupancies. For a target dopant count $N_t$, with $N_s$ eligible
	candidate sites in the shell and $N_b$ in the bulk, the selection
	probabilities are
	\begin{equation}
		p_{\mathrm{bulk}}=\frac{N_t}{N_b+E N_s},
		\qquad
		p_{\mathrm{shell}}=E\,p_{\mathrm{bulk}}.
		\label{eq:doping-probabilities}
	\end{equation}
	The shell contains sites whose boundary-margin measure, supplied by
	the tessellation backend, does not exceed \code{shell\_width}.
	This parameter denotes the shell half-thickness. A required
	probability above unity is rejected as infeasible. Here, $E=1$
	gives equal selection probabilities in both regions, $E>1$ favours
	the shell, and $E\to0$ suppresses shell occupation.
	
	After placement, the dopant atom fraction and achieved enrichment are
	evaluated separately:
	\begin{equation}
		x_{\mathrm{dop}}=\frac{N_{\mathrm{dop}}}{N_{\mathrm{final}}},
		\qquad
		E_{\mathrm{achieved}}
		=
		\frac{n_{\mathrm{dop},s}/N_s}
		{n_{\mathrm{dop},b}/N_b}.
		\label{eq:doping-achieved}
	\end{equation}
	Thus, the local site fraction reported in \code{doping\_profile.csv}
	is the number of placed dopants divided by the number of eligible
	candidate sites in each distance bin; it is not the dopant fraction
	among all atoms in that bin. Per-grain placement is recorded in
	\code{doping.csv}.
	
	Bernoulli sampling produces fluctuations around the target counts.
	For interstitial insertion, subsequent pruning of conflicts among
	selected sites can further reduce the achieved concentration without
	re-solving the selection probabilities. Gate G17 warns when the final
	dopant atom fraction differs from its target by more than one
	percentage point, or when enrichment differs by more than 15\%
	relative or cannot be evaluated. Enrichment estimates become less
	reliable as the bulk dopant population decreases. Gate G18 independently
	re-verifies the minimum-distance constraint for interstitial dopants
	using a fresh periodic neighbour query. Gates G8/G9 assess the
	pre-doping host snapshot.

	The shell/bulk model is a deterministic geometric placement scheme, not a
	thermodynamic model: segregation energies, temperatures, and chemical
	potentials do not enter, and $E$ is a prescribed input rather than an
	equilibrium prediction. Site-specific equilibrium segregation is therefore
	left to downstream MD or Monte Carlo. The \code{doping} examples cover substitutional replacement and
	interstitial insertion, with interstitial sites specified through
	named presets or explicit coordinates expanded into symmetry orbits,
	including for CIF-derived hosts (\ref{app:examples}).
	
	Fig.~\ref{fig:doping} demonstrates the machinery on anatase TiO$_2$
	(CIF-sourced, $I4_1/amd$ detected by spglib from a P1 export): 2~at\% Li
	interstitials on the Wyckoff-$4b$ octahedral (intercalation) sublattice, with
	GB-targeted dopant placement in a shell of 3~\AA{} half-thickness. Panels (a,b) compare two
	enrichment strengths on the same 6-grain host: moderate enrichment ($E=20$,
	with a visible bulk population) and strong enrichment ($E=100$, with $28\,052$
	of $29\,959$ Li atoms, or $93.6\%$, in the GB shell). Panel (c) reports the achieved local Li site fraction for
	$E\in\{10^{-5},20,40,60,80,100\}$ at a common target concentration
	of 2~at\% in the final structure. For $E=20$--$100$, the same
	6-grain host is used: the shell-averaged site fraction increases
	from approximately $0.35$ to $0.44$, while the bulk site fraction
	decreases from $0.018$ to $0.0045$. The $E=10^{-5}$ case illustrates
	the shell-depletion limit on a separate 10-grain host, with no shell
	dopants in the reported realisation and a bulk site fraction of
	$0.075$. It is therefore an illustrative limiting case rather than
	a controlled extension of the same-host enrichment series.
	The no-preference condition is $E=1$, which is not included in
	this panel. Panel (d) shows a two-phase anatase/rutile polycrystal with a 50/50
	target, for which gate G15 measures $0.503/0.497$, built with the same
	curved-boundary construction. It provides a starting structure for studies of
	GB- and interphase-dominated transport.

	\subsection{Multiphase and single crystal}
	\label{sec:multiphase}
	
	\code{phases:} assigns each grain to a phase using a deterministic greedy
	(longest-processing-time) partition of the measured pre-fill grain volumes.
	The longest-processing-time rule is heuristic and is not guaranteed to yield a
	globally optimal partition. Following the metallographic convention, the
	achieved volume fractions approximate the requested target values subject to
	grain-count granularity. Gate G15 reports the resulting deviation; for the
	50/50 anatase/rutile system in Fig.~\ref{fig:doping}(d), for example, it
	measures $0.503/0.497$. Phases may differ in space group, lattice parameters,
	and elemental composition. Interphase boundaries carry an undefined
	misorientation but retain their habit-plane Miller indices.
	
	Multiphase outputs are geometric starting structures, not relaxed interfaces,
	and require MD energy minimisation and equilibration before use.
	\grainsmith{} imposes no crystallographic orientation relationship
	(Kurdjumov--Sachs, Nishiyama--Wassermann, or Burgers) across phases: each phase
	is oriented independently, and the two lattices meet on the Voronoi habit
	plane. With differing lattice parameters and densities, the resulting
	interphase boundary is therefore generally high-energy and far from
	equilibrium. The uniform $0.85\,d_{nn}$ overlap removal only deletes
	geometrically coincident atoms and does not construct a coherent interface.
	Imposing interphase orientation relationships is planned future work
	(Sec.~\ref{sec:conclusions}).
	
	Gate G15 checks phase volume fractions against their targets
	using the granularity scale $V_{\max}/V_{\mathrm{box}}$,
	where $V_{\max}$ is the largest grain volume. This bounds
	the fraction change from reassigning one grain, not the
	final assignment error: exact agreement is possible, while
	phase-occupancy constraints can produce larger deviations.
	
	For a single crystal (one grain), gate G14 checks box/lattice commensurability using $\varepsilon=\lVert(R\!\cdot\!A)(m-\mathrm{round}\,m)\rVert/L \le 10^{-8}$, where $A$ is the unrotated lattice matrix (columns are lattice vectors), $R$ is the grain orientation, $m$ is the fractional-coordinate solution for the box vector in the rotated cell, and $L$ is that box vector's length. Interface areas are written $A_{ij}$ and $A_{\mathrm{GB}}$ throughout and are always subscripted.
	
	An axis-aligned box that is incommensurate with the oriented lattice
	can introduce artificial self-boundary defects at periodic joins.
	The G14 residual measures geometric lattice mismatch, not an imposed
	homogeneous elastic strain. Exact commensurate orthogonal supercells
	are available only for compatible lattice parameters and orientations. For fully periodic single crystals, \code{box.cells:} $[n_1,n_2,n_3]$
	constructs the box vectors as integer multiples of the conventional-cell
	vectors. This path requires a fixed identity orientation and zero vacuum,
	so the box and reference lattice are aligned by construction.
	G14 checks the resulting residual lattice mismatch against the
	$10^{-8}$ tolerance. The resolved LAMMPS restricted-triclinic box,
	including automatic integer lattice-vector reduction when a raw tilt factor
	exceeds the LAMMPS bound, is echoed in the provenance outputs. The PBC overlap
	search and gate G7 use a general triclinic minimum-image treatment with a
	27-image ghost replication when the cutoff satisfies the cell's
	reciprocal-norm bound on every periodic axis, together with a
	reciprocal-norm-bounded image search for more skewed cells; per-axis
	fractional rounding is not a minimum-image algorithm in a tilted cell. Gate
	G10 re-verifies the emitted \code{xy xz yz} tilt line and atom containment.
	
	The polycrystal tessellation core remains restricted to orthogonal boxes
	and does not impose a common lattice-commensurability condition on all
	grains. The \code{owns()} home-cell rule of
	Sec.~\ref{sec:tessellation} provides unique geometric ownership within
	the periodic domain, while overlap removal and G7 enforce the
	minimum-distance constraint across periodic images. These operations
	do not independently certify crystallographic continuity across a
	periodic join. In particular, where a grain meets its own periodic
	image, a non-lattice translation can introduce an artificial join
	defect even if the minimum-distance check passes.
	
	G14 explicitly assesses commensurability on the single-crystal path.
	It reports excessive mismatch as a warning, allowing configurations
	with intentional periodic join defects to be retained. A
	non-commensurate box still defines a periodic simulation domain,
	but should not be interpreted as a defect-free periodic crystal
	solely because its interatomic-distance checks pass. The \code{alloys\_multiphase} group ships two-phase composites differing in lattice
	and element, as well as a random solid solution. Single-crystal examples
	range from a cubic box to an exactly commensurate triclinic cell in the
	\code{basics} and \code{cif} groups (\ref{app:examples}).
	
	\subsection{Analysis}
	\label{sec:analysis}
	The disorientation between two grains is defined as the minimum-angle
	misorientation over the double coset of the point group, with symmetry applied
	on both sides; it is therefore symmetry-invariant by construction.
	\grainsmith{} reports both the disorientation angle and its rotation axis. A
	cubic coincidence-site-lattice (CSL) assignment labels a boundary as
	$\Sigma N$ when its disorientation falls within the Brandon window
	$\Delta\theta\le15^\circ/\sqrt{N}$ of an exact CSL rotation
	\cite{brandon1966csl}. Each $(\Sigma,\theta,\langle uvw\rangle)$ table entry
	is verified as a genuine member of its misorientation class, with
	double-coset deviation $\approx0$ from the standard reference. This
	classification is strictly geometric: a $\Sigma$ label does not by itself
	imply a reduced boundary energy or any other special physical property.
	
	This analysis provides statistical classification across the generated
	boundary network and is distinct from the atomic CSL construction performed
	by dedicated bicrystal tools such as aimsgb~\cite{cheng2018aimsgb} and
	GB\_code~\cite{hadian2018gbcode}, which construct a single O-lattice boundary.
	\grainsmith{} labels boundaries; it does not construct coincident bicrystals.
	Boundary character is determined from the physical minimal-angle map between
	the boundary-plane normal and the disorientation axis. With $\psi$ denoting
	the angle between them, a boundary is classified as twist when
	$\psi<15^\circ$ (\code{CHAR\_TWIST\_DEG}), tilt when
	$\psi>75^\circ$ (\code{CHAR\_TILT\_DEG}), and mixed otherwise. Boundaries
	lacking a defined disorientation axis are designated as undefined, while
	cross-phase boundaries are reported as interphase. Per-class area fractions
	are compiled into the boundary statistics.
	
	Metric properties derived from voxel representations require distinct
	treatment. The label-change voxel-face (staircase) estimator used for
	boundary area, sphericity, and metric length density $L_V$ over-counts
	grid-oblique boundary area by the factor $\lVert\hat{n}\rVert_1$, reaching a
	maximum bias of $\sqrt{3}$ for the body-diagonal $(111)$ orientation relative
	to the Cartesian grid. This bias is a property of the staircase estimator,
	not of voxel representations in general; marching-surface and Crofton-type
	estimators on the same grid exhibit different scaling behaviour. Consequently,
	flat, power, and single-crystal geometries use explicitly reconstructed
	polyhedral-cell estimators, and only curved or imported geometries carry this
	staircase bias, recorded in the estimator field of \code{statistics.csv}. The
	log-normal goodness-of-fit $p$-value evaluates a fitted distribution and is
	therefore optimistic because the parameters are estimated from the same data,
	subject to the standard Lilliefors caveat~\cite{lilliefors1967ks}.
	
	An opt-in module (\code{analysis.gb\_curvature}, default off) evaluates the
	local grain-boundary mean and Gaussian curvatures $(H,K)$ at discrete sample
	points. The calculation uses central differences on the implicit level set
	of the margin difference together with standard implicit-surface curvature
	formulations~\cite{goldman2005curvature}. The numerical kernel is validated
	against analytic spheres, cylinders, and planes and cross-checked against
	marching-cubes mesh reconstructions. Area-weighted aggregate statistics are
	recorded in \code{boundaries.csv}, whereas local samples are exported to
	\code{gb\_curvature.csv}; flat geometries yield $H=K=0$ identically. For
	curved and imported geometries, area-weighted aggregates inherit the
	staircase boundary-area bias described above, whereas local $H$ and $K$ are
	evaluated directly on the level set and do not inherit this area-estimator
	bias. Curvature magnitudes approaching $1/(2h)$, where $h$ is the grid
	spacing, cannot be resolved by this finite-difference stencil, representing
	a practical resolution limit set by the spatial discretization. Separately,
	sample points with level-set gradient magnitudes below
	\code{CURV\_GRAD\_MIN} are degenerate for the curvature formulation and are
	pruned; their relative fraction is reported by the warn-only gate G16.
	
	Gate G21 monitors the integrated Gaussian curvature over the retained
	face-interior samples of each grain:
	\begin{equation}
		R_{\mathrm{GB}}^{(i)}
		=
		\Bigl\lvert
		\sum_{s\in\mathcal{F}_i} K_s\,A_s
		\Bigr\rvert,
		\label{eq:g21}
	\end{equation}
	where $\mathcal{F}_i$ is the sampling set for grain $i$ and $A_s$ is the
	area weight assigned to sample $s$. For a complete closed orientable
	surface of genus $g$, the Gauss--Bonnet theorem gives
	$\oint_S K\,dA=4\pi(1-g)$. This identity does not, however, impose a
	$4\pi$ upper bound on Eq.~\eqref{eq:g21}, which excludes parts of the
	grain surface and uses discretised area weights. For an ideal
	polyhedron, the intrinsic Gaussian curvature is concentrated at the
	vertices as angle defects, while planar face interiors contribute zero;
	this observation does not establish a corresponding bound for the
	curved surfaces considered here.
	
	The implementation issues a warning when
	$R_{\mathrm{GB}}^{(i)}>4\pi$ for any grain. The threshold is a heuristic
	reference scale, motivated by the total curvature of a closed genus-zero
	surface, rather than a theorem-based acceptance criterion. The sampled
	integral can reflect genuine face curvature, cancellation between
	positive and negative contributions, omitted surface regions, and
	numerical errors in curvature estimation and area weighting. Neither
	exceeding the threshold nor remaining below it establishes the accuracy
	of the curvature estimator or the physical validity of the atomic
	configuration.
	
	The documented calibration found worst-grain values of
	$0.977$--$1.041$ times the threshold for near-flat cases and approximately
	$5$--$14$ times the threshold for ordinary curved cases. Refining the
	voxel grid from 32 to 128 produced a non-monotonic response without a
	clear downward trend. A two-grain periodic bicrystal without triple
	junctions also exceeded the threshold, so triple-junction sampling alone
	does not explain the observed behaviour. These tests do not establish
	convergence or identify a unique error mechanism. G21 is therefore
	retained as a warning about the quantitative reliability of the
	curvature estimates, separate from the geometric and interatomic-distance
	checks.
	
	% =====================================================================
	\section{Validation}
	\label{sec:validation}
	Assessment comprises two complementary stages. First, construction
	verification checks geometric identities and constraints and quantifies
	deviations from prescribed statistical targets. Second, representative
	MD simulations examine the relaxation and short-time structural
	stability of selected configurations under a specified interatomic
	potential. These simulations assess behaviour beyond the construction
	checks, with conclusions restricted to the tested systems and protocol.
	
	\subsection{Structural and statistical re-measurement}
	\label{sec:validation-structural}
\begin{figure*}[!t]
		\centering
		\begin{subfigure}[t]{0.29\linewidth}
			\includegraphics[width=\linewidth]{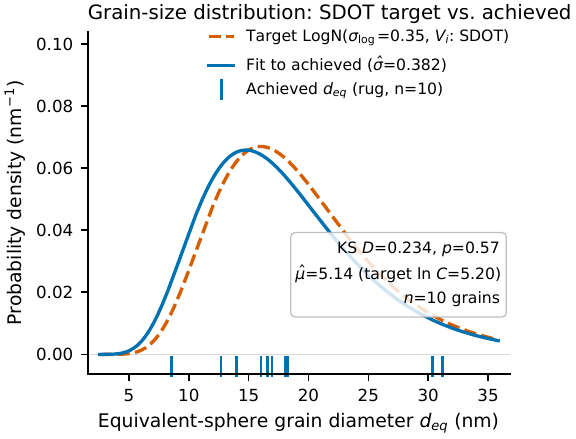}
			\caption{}\label{fig:validation-a}
		\end{subfigure}\hspace{0.4cm}
		\begin{subfigure}[t]{0.29\linewidth}
			\includegraphics[width=\linewidth]{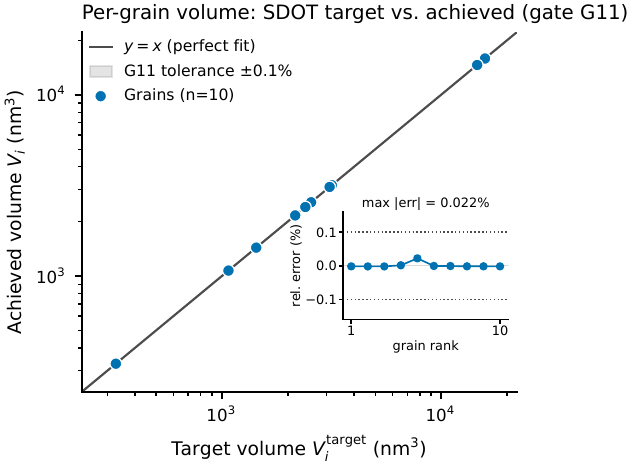}
			\caption{}\label{fig:validation-b}
		\end{subfigure}\\[2pt]
		\begin{subfigure}[t]{0.29\linewidth}
			\includegraphics[width=\linewidth]{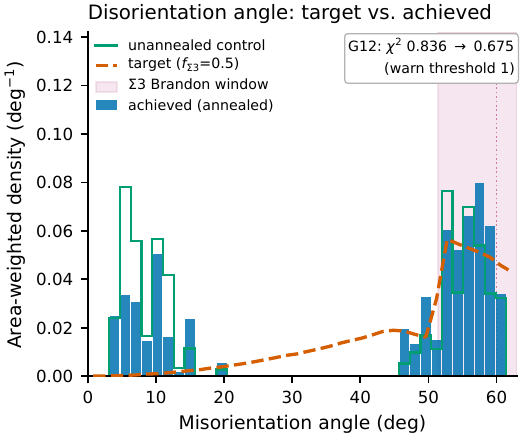}
			\caption{}\label{fig:validation-c}
		\end{subfigure}\hspace{0.4cm}
		\begin{subfigure}[t]{0.29\linewidth}
			\includegraphics[width=\linewidth]{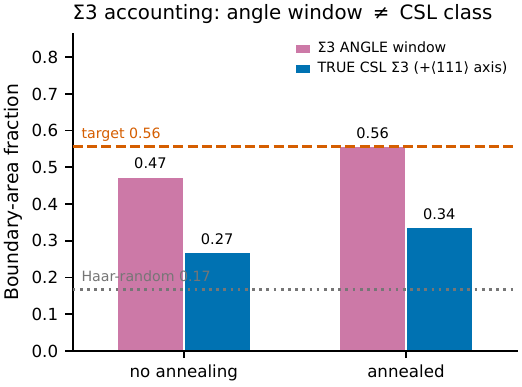}
			\caption{}\label{fig:validation-d}
		\end{subfigure}
		\caption{Structural and statistical validation of the two representative runs.
			(a)~Target log-normal grain-size law ($\sigma_{\log}=0.35$, dashed) and
			refitted distribution of the 10 achieved equivalent-sphere diameters
			($\hat{\sigma}=0.382$, solid; rug marks: individual grains) for the
			$(360~\text{\AA})^3$ FCC Cu run. (b)~Achieved versus target per-grain volume
			on the same run (log-log), with the $y=x$ line and relative-error inset
			against the G11 tolerance of $\pm0.1\%$. (c)~Boundary-area-weighted
			disorientation-angle distribution of the twin-textured polycrystal (bars),
			with the $\Sigma3$-enriched target (dashed), unannealed control at the same
			seed and ODF (step curve), shaded $\Sigma3$ Brandon window
			($60^\circ\pm15^\circ/\sqrt{3}$), and the G12 verdict annotation.
			(d)~Boundary-area fraction within the angular window versus
			the CSL $\Sigma3$ area fraction before and after annealing.
			The target and Haar-random reference levels are histogram-mass
			sums over bins whose centres lie within the angular window.}
		\label{fig:validation}
	\end{figure*}

	This stage tests the generator against its principal quantitative contracts:
	properties specified as construction targets or constraints are independently
	re-measured on the generated structure and compared with their configured
	values. For properties subject to stochastic sampling or optimization bounds,
	the framework reports the quantified deviation or diagnostic status.
	Fig.~\ref{fig:validation} demonstrates this verification loop for two
	representative runs: an FCC Cu polycrystal (10 grains, $(360~\text{\AA})^3$,
	$\sim$3.9M atoms, seed 360360) with a log-normal equivalent-diameter target
	($\sigma_{\log}=0.35$), and a twin-textured Cu polycrystal (14 grains,
	$(60~\text{\AA})^3$, with two ODF components related by the $\Sigma3$ rotation
	at equal volume weight, \code{sigma3\_angle\_enriched} target with
	$f_{\Sigma3}=0.5$, seed 400412).
	
	Fig.~\ref{fig:validation}(a) compares the realised grain-size distribution with
	the prescribed log-normal target. A log-normal refit of the 10 achieved
	equivalent-sphere diameters (gate-independent and recorded in
	\code{statistics.csv}) gives $\hat{\sigma}=0.382$ for the prescribed
	$\sigma_{\log}=0.35$; the corresponding Kolmogorov--Smirnov test yields
	$D=0.234$ and $p=0.57$ against the target law ($n=10$). At this small sample
	size, the KS test has limited discriminating power, so non-rejection is
	interpreted as consistency with the target rather than evidence of
	distributional agreement. The effect of finite grain count on the realised
	shape parameter is quantified by the replicated sweep in Table~\ref{tab:b2}.
	
	\begin{table}[!htb]
		\centering
		\caption{Log-normal grain-size targeting versus realised distribution shape
			for the replicated sweep (target $\sigma_{\log}=0.35$; G11 passes in every replicate).
			$N$: grain count; box: cubic box edge at constant mean grain volume;
			reps: independent seeds per $N$; $\hat{\sigma}$: mean $\pm$ standard deviation
			across replicates of the log-normal shape parameter refitted to the achieved
			equivalent-sphere diameters; s.e.\ theory: expected sampling error
			$\sigma_{\log}/\sqrt{2(N-1)}$ for an ideal log-normal $N$-sample.}
		\label{tab:b2}
		\footnotesize
		\setlength{\tabcolsep}{4.5pt}
		\begin{tabular}{@{}rrrcc@{}}
			\toprule
			$N$ & box (\AA) & reps & $\hat\sigma$ (mean $\pm$ sd) & s.e.\ theory \\
			\midrule
			10   & 56.5  & 50 & $0.356\pm0.085$ & 0.082 \\
			100  & 121.6 & 25 & $0.358\pm0.025$ & 0.025 \\
			1000 & 262.1 & 10 & $0.350\pm0.005$ & 0.008 \\
			\bottomrule
		\end{tabular}
	\end{table}
	
	Fig.~\ref{fig:validation}(b) compares achieved and target per-grain volumes.
	The SDOT damped-Newton fit converges in 4 iterations, and all 10 achieved
	volumes lie on the $y=x$ line within a maximum relative error of
	$2.2\times10^{-4}$, well inside the G11 tolerance of $10^{-3}$. The per-grain
	targets are regenerated deterministically from the recorded seed through the
	\code{sizes} RNG stream, and the re-measured errors agree with
	\code{summary.csv} to all printed digits. Across grain counts
	($N\in\{10,100,1000\}$), the SDOT damped-Newton fit converges in 3--8 iterations
	and satisfies G11 in every replicate. The realised distribution shape behaves
	as a finite-sample statistic: the mean fitted $\hat{\sigma}$ remains centred on
	the target $\sigma_{\log}=0.35$, while its spread across replicates matches the
	expected theoretical sampling error (Table~\ref{tab:b2}). Thus, deviations of
	$\hat{\sigma}$ at small grain counts reflect finite-sample variation rather than
	a systematic fitting error.
	
	Figs.~\ref{fig:validation}(c,d) assess shaping of the
	boundary-area-weighted disorientation-angle distribution.
	Assignment annealing (20\,000 steps, 63 accepted swaps) increases
	the area fraction within the $\Sigma3$ Brandon angular window
	from 0.474 to 0.558 relative to the unannealed reference with the
	same seed and ODF. For these configurations, the angular-window fractions agree
	with histogram sums over bins whose centres lie within the
	window. The corresponding target and Haar-random sums are
	0.556 and 0.166; these bin-centre sums are distinct from
	integration over the exact window limits. Despite the close
	agreement of the achieved and target sums, the full-histogram
	discrepancy remains appreciable, with the symmetric $\chi^2$
	decreasing from 0.836 to 0.675. The two $\Sigma3$ measures
	are not interchangeable: the Brandon-window mass is angle-only, whereas the
	CSL $\Sigma3$ area fraction of Sec.~\ref{sec:texture} also requires the
	$\langle111\rangle$ axis test and is therefore smaller, increasing from 0.268
	to 0.337 for the same runs. The latter is recorded in \code{statistics.csv} and
	reported alongside the angle-window measure by G23. The conditional probability
	of the $\langle111\rangle$ axis test given window membership increases from
	0.566 to 0.604. Fig.~\ref{fig:validation}(d) shows both measures before and
	after annealing: the permutation raises both, while the gap persists because
	the axis test is not part of the annealing objective. In panel~(c), the shaded
	Brandon window is clipped at the $63^\circ$ upper edge of the bin grid.
	
	Because grain volumes are unequal, the permutation does not conserve the
	volume-weighted ODF (Sec.~\ref{sec:texture}). The run therefore sets \code{odf\_drift\_max}${}=0.03$ for the discrete total-variation distance in Eq.~\eqref{eq:odf-drift}. The annealer vetoed
	14\,948 candidate swaps that would have breached this cap. Using the
	final orientation assignment and tessellation grain volumes, G22
	independently re-measures a drift of 0.0288 relative to the initial
	assignment. The number-weighted orientation set is unchanged by construction,
	while G26 separately reports the discrepancy between tessellation-volume
	and atom-count grain weights defined in Eq.~\eqref{eq:weighting-gap}. For this example, \code{chi2\_max}=1 is used instead of the default
	G12 warning threshold of 0.5 (\ref{app:gates}). The final
	$\chi^2=0.675$ therefore remains below the configured threshold but
	exceeds the default. This setting changes the warning criterion,
	not the achieved histogram or its discrepancy from the target.
	The two-component ODF, centred on the cube orientation and its
	$\Sigma3$ twin, favours same-component low-angle and cross-component
	near-$60^\circ$ pairs. It therefore restricts the population available
	to reproduce the Haar-random-shaped intermediate-angle part of the
	target. The residual is consistent with these restrictions, although
	the finite annealing run alone does not establish the minimum
	attainable $\chi^2$. The observed increase in angular-window mass
	demonstrates redistribution toward the target window rather than
	agreement with the full target histogram.
	
	The angle-shaping mechanism exhibits a measurable, target-dependent shortfall.
	A dedicated campaign quantifies this using 40 grains in a $(100~\text{\AA})^3$
	FCC Cu box with Haar-random orientations, a \code{sigma3\_angle\_enriched}
	target swept over $f_{\Sigma3}=0.10$--$0.50$, three seeds per target, and
	20\,000 annealing steps per run
	(\code{benchmarks/validation/}\allowbreak\code{validate\_mdf.py}, recorded in
	\code{mdf\_campaign/}\allowbreak\code{mdf\_validation.csv}). Across all fifteen
	runs, the annealer's self-reported $\chi^2$ agrees with the independently
	re-measured G12 value to full printed precision.
	
	At the fixed annealing budget of 20\,000 steps, the relative
	shortfall from the target window mass increases with
	$f_{\Sigma3}$. The achieved area-weighted Brandon-window mass
	reaches $0.91$--$0.97$ of its target at $f_{\Sigma3}=0.10$,
	but only $0.74$--$0.75$ at $0.50$. Correspondingly, the G12
	$\chi^2$ increases monotonically with $f_{\Sigma3}$ for each
	of the three seeds, from $0.011$--$0.027$ to
	$0.104$--$0.126$, while remaining below the default warning
	threshold of 0.5. A Haar-random ODF can supply most of what a Mackenzie-based target
	requires, giving a small residual, whereas the two-component twin ODF of
	Fig.~\ref{fig:validation}(c) yields a residual $\chi^2$ of 0.675, several-fold
	above the largest value in this sweep.
	
	Across the sweep, the CSL $\Sigma3$ area fraction, which additionally requires
	the $\langle111\rangle$ axis condition, remains at $0.021$--$0.060$, while
	the Brandon-window area fraction spans $0.241$--$0.442$. Populating the
	disorientation-angle window therefore does not by itself create $\Sigma3$
	boundaries, so G23 reports the two quantities separately. Assignment annealing
	only re-assigns existing orientations to grains, leaving the number-weighted
	orientation set unchanged; the attainable $\Sigma3$ content is consequently
	bounded by the twin content of the fixed ODF and the Voronoi adjacency graph.
	G12 reports the discrepancy from the configured target as a warning
	diagnostic. It does not determine the optimum attainable by permutation
	or separate restrictions imposed by the orientation set and graph
	from incomplete optimisation.
	
	\begin{figure}[!htb]
		\centering
		\includegraphics[width=0.8\linewidth]{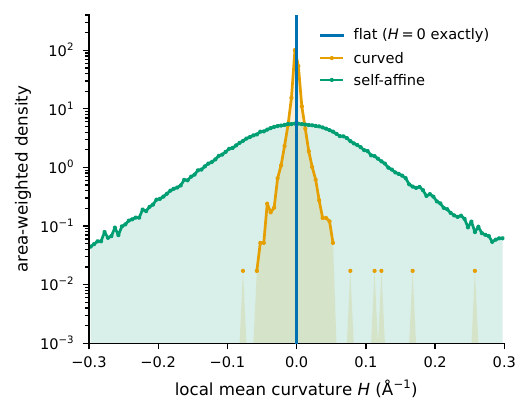}
		\caption{Area-weighted distribution of local grain-boundary mean curvature $H$
			for the three boundary geometries of Fig.~\ref{fig:geometries}(a--c) in the
			10-grain FCC Cu system: planar baseline at $H=0$ (vertical line), curved
			backend ($|H|\lesssim0.05$~\AA$^{-1}$), and self-affine spectrum
			($\mathcal{H}=0.8$). Isolated dots indicate sparsely occupied bins.}
		\label{fig:curvature}
	\end{figure}

	The boundary-morphology families are characterised by the opt-in
	local-curvature analysis. Fig.~\ref{fig:curvature} compares the
	area-weighted distribution of local mean curvature $H$ across the three
	grain-boundary geometry families of Fig.~\ref{fig:geometries}(a--c), measured
	at boundary sample points from the same 10-grain FCC Cu system. Flat faces
	report $H=0$ exactly, providing the analytic planar limit of the curvature
	kernel. The anisotropic curved backend produces a narrow distribution within
	$|H|\lesssim0.05$~\AA$^{-1}$, consistent with gentle grain-scale face bowing,
	whereas the self-affine construction ($\mathcal{H}=0.8$) broadens the
	distribution by approximately an order of magnitude, with heavy symmetric
	tails reflecting the band-limited roughness spectrum. The three constructions
	therefore produce distinct realised curvature distributions, subject to the
	area-weighting bias of the boundary-area estimator; no formal hypothesis test
	is attached to this comparison. The self-affine run carries a G21 warning, with a maximum per-grain value of $R_{\mathrm{GB}}^{(i)}=350.6$, compared with the heuristic
	threshold of $4\pi\approx12.57$, and a degenerate-sample fraction of
	$0.001\%$. The small discarded fraction indicates that few samples
	violate the gradient criterion, but does not establish the accuracy of
	the retained curvature estimates. Likewise, the large G21 value does
	not by itself distinguish genuine integrated face curvature from
	sampling and discretisation effects. Figure~\ref{fig:curvature} therefore
	provides a descriptive comparison at the chosen numerical resolution;
	the absolute curvature values and quantitative differences between
	geometry families require a resolution-convergence assessment.
	
	Finally, the voxel bridge is verified by a round-trip on the shipped synthetic
	label field (8 grains, $32^3$ voxels;
	\cfg{examples/voxel_import/fe_voxel_import.yaml}) atomised to BCC Fe, yielding
	a per-grain volume-fraction round-trip error of 0.0 to floating-point precision
	and a grain-adjacency Jaccard index of 1.0 (28/28 contacts). The repository
	test suite additionally asserts exact per-grain volume and adjacency agreement
	for a voxelised Voronoi round-trip.

	\subsection{Stability under a representative MD protocol}
	\label{sec:validation-md}
	Construction checks establish geometric consistency and quantify
	agreement with the requested microstructure, but do not determine its
	response to an interatomic potential. Tessellation-based assembly and
	overlap removal can leave interfaces that reconstruct during relaxation
	or subsequent dynamics. We therefore examine whether the generated
	polycrystals retain crystalline interiors and their grain-scale
	structure under a representative MD protocol. Testing smaller cells
	at fixed grain count additionally probes configurations with a larger
	boundary area per atom.
	
	The three boundary-geometry families of Fig.~\ref{fig:geometries}(a--c) were
	evaluated dynamically using LAMMPS~\cite{thompson2022lammps} with the Mishin
	et al.\ EAM Cu potential~\cite{mishin2001cu} and a 2~fs timestep. Each geometry
	was tested at $(360~\text{\AA})^3$ (3.86--3.88M atoms) and
	$(160~\text{\AA})^3$ (0.33M atoms) using the same 10-grain configuration
	(seed 282930). At fixed grain count, the smaller box increases the boundary
	area per atom by a factor of 2.27--2.31 (Table~\ref{tab:md-gb-energy}),
	providing a deliberately stringent stability test across the same
	seed-defined grain ensemble. The protocol comprises conjugate-gradient
	position minimisation, joint position-and-cell minimisation, 100~ps NPT
	equilibration at 300~K and zero pressure, 200~ps NPT production, and a final
	re-quench to a 0~K minimum. Throughout this section, ``as-built'' denotes the atomic configuration
	exported by \grainsmith{} before any energy minimisation or MD treatment. For the six polycrystals, the time-averaged production temperatures
	deviated from 300~K by at most 0.04~K. The reported residual pressure
	magnitudes were below 0.05~bar. For the two perfect-crystal reference
	runs used to establish the 300~K baseline, the corresponding reported
	temperature deviations and pressure magnitudes were within 0.24~K
	and 0.38~bar, respectively.
	
\begin{figure}[!tb]
		\centering
		\begin{subfigure}[t]{\linewidth}
			\includegraphics[width=0.95\linewidth]{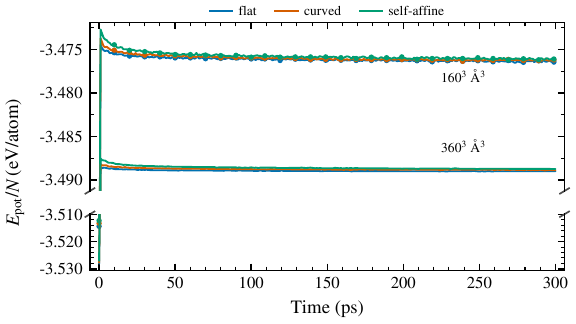}
			\caption{}\label{fig:md-stability-a}
		\end{subfigure}\\[3pt]
		\begin{subfigure}[t]{\linewidth}
			\includegraphics[width=\linewidth]{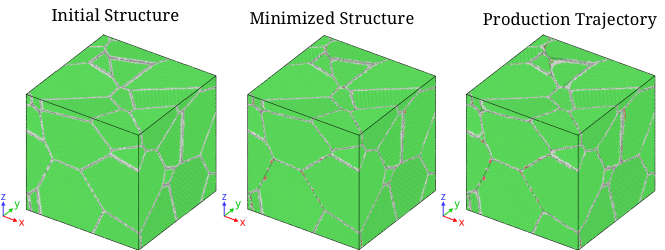}
			\caption{}\label{fig:md-stability-b}
		\end{subfigure}\\[2pt]
		\begin{subfigure}[t]{\linewidth}
			\includegraphics[width=\linewidth]{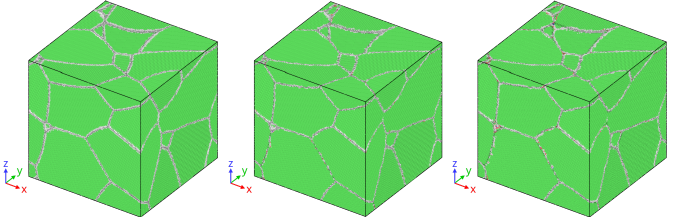}
			\caption{}\label{fig:md-stability-c}
		\end{subfigure}\\[2pt]
		\begin{subfigure}[t]{\linewidth}
			\includegraphics[width=\linewidth]{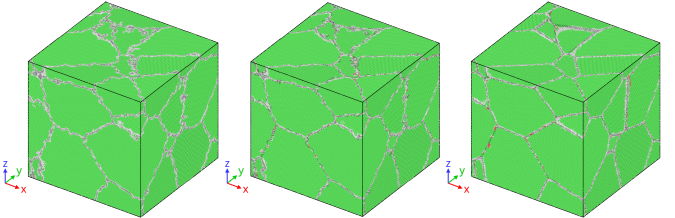}
			\caption{}\label{fig:md-stability-d}
		\end{subfigure}
		\caption{Molecular-dynamics stability of the boundary-geometry families.
			(a)~Trajectory potential energy per atom for all six models (broken
			ordinate separates 0~K minima from 300~K plateaus). (b--d)~PTM
			structural snapshots for the $(360~\text{\AA})^3$ models: (b)~flat,
			(c)~curved, and (d)~self-affine boundaries. Columns left to right:
			as-built, 0~K minimum, and 300~K production. Green: FCC; white:
			unclassified boundary atoms; red: HCP.}
		\label{fig:md-stability}
	\end{figure}

	Fig.~\ref{fig:md-stability}(a) displays the trajectory potential energy per
	atom. Each curve increases by 38.4--39.5~meV/atom during thermalization, measured
	from the 0~K minimum to $t=5$~ps, followed by a further 0.34--2.05~meV/atom
	relaxation from that point to the 300~K block average over the remaining
	295~ps. Following the initial transient, the plotted potential energies approach
	plateaus without abrupt drops on the displayed time scale. This
	observation characterises the potential-energy evolution under NPT
	conditions; it is not an energy-conservation test. The reported 300~K plateau energies differ only slightly among the
	three initial boundary morphologies at each box size:
	$-3.4890$ to $-3.4887$~eV/atom at $(360~\text{\AA})^3$ and
	$-3.4763$ to $-3.4760$~eV/atom at $(160~\text{\AA})^3$. Relative to the
	perfect-crystal reference ($-3.5004$~eV/atom at 300~K), the excess energy is
	11.4--11.7~meV/atom in the large cells and 24.1--24.5~meV/atom in the small
	cells. This 2.09--2.12-fold increase accompanies the 2.27--2.31-fold increase
	in boundary area per atom, indicating that the higher energy in the smaller
	cells is consistent with the increased boundary contribution rather than a
	continuing instability.
	
	Fig.~\ref{fig:md-stability}(b--d) shows structural snapshots classified
	by polyhedral template matching (PTM)~\cite{larsen2016ptm} in
	OVITO~, using an RMSD threshold of 0.1.
	In the displayed configurations, grain interiors remain predominantly
	FCC, while non-FCC atoms are concentrated near the boundary network.
	The snapshots show no apparent widespread amorphisation or loss of
	the grain-scale structure. The initial and minimised configurations
	retain similar boundary morphologies at the displayed scale, although
	this visual comparison does not quantify atomic displacements or the
	energy released during minimisation. During the 300~K stage, the
	curved-boundary example retains visible grain-scale bowing, whereas
	the finer asperities of the self-affine example become less pronounced,
	giving a morphology closer to the flat reference. These observations
	distinguish retention of the grain-scale structure from preservation
	of the imposed fine-scale boundary roughness.

	The excess potential energies of the initial 0~K minimum and the final
	re-quenched minimum provide a complementary quantitative
	comparison. Table~\ref{tab:md-gb-energy} reports
	\begin{equation}
		\gamma_{\mathrm{ex}} = \frac{E_{\mathrm{tot}}-N_{\mathrm{at}}E_{\mathrm{coh}}}{A_{\mathrm{GB}}},
		\label{eq:gamma-ex}
	\end{equation}
	where $E_{\mathrm{tot}}$ is the total potential energy of the corresponding
	0~K minimum, $N_{\mathrm{at}}$ the atom count after overlap removal,
	$E_{\mathrm{coh}}=-3.5400$~eV/atom, and $A_{\mathrm{GB}}$ the as-built
	boundary area recorded by the generator in \code{boundaries.csv}. Here, $\gamma_{\mathrm{ex}}$ measures the total excess potential
	energy per unit as-built boundary area, including contributions
	from triple junctions and residual defects. It therefore provides
	a common normalisation for the before-and-after comparison,
	rather than a direct measure of the relaxed grain-boundary energy. The 300~K excursion and the final re-quench reduce
	$\gamma_{\mathrm{ex}}$ relative to the initial 0~K minimum
	in every case ($\Delta\gamma_{\mathrm{ex}}=-42.5$ to $-62.3$~mJ/m$^2$). The variation across
	morphologies reflects differences in both the excess energy and the as-built
	area denominator and is therefore not interpreted as an intrinsic
	morphology-dependent boundary energy.
	
	\begin{table}[!htb]
		\centering
		\caption{Area-normalised excess potential energy
			$\gamma_{\mathrm{ex}}$ for the six MD runs.
			Initial 0~K min refers to the minimum obtained after the
			position and cell minimisation stages, before the 300~K stage;
			annealed refers to the final re-quenched minimum after the
			full relaxation protocol. Both values use the same fixed
			as-built boundary area $A_{\mathrm{GB}}$ from \code{boundaries.csv}.
			Thus, $\Delta\gamma_{\mathrm{ex}}$ measures the change over the 300~K
			excursion and re-quench, while comparisons between geometries also depend
			on their different area denominators.}
		\label{tab:md-gb-energy}
		\footnotesize
		\setlength{\tabcolsep}{4.5pt}
		\begin{tabular}{@{}llcccc@{}}
			\toprule
			box (\AA) & geometry & $A_{\mathrm{GB}}/N$ (\AA$^2$)
			& \multicolumn{2}{c}{$\gamma_{\mathrm{ex}}$} (J/m$^2$) & $\Delta\gamma_{\mathrm{ex}}$ \\
			\cmidrule(l){4-5}
			& & & initial 0~K min & annealed & (mJ/m$^2$) \\
			\midrule
			$360^3$ & flat        & 0.215 & 0.8815 & 0.8390 & $-42.5$ \\
			$360^3$ & curved      & 0.319 & 0.6165 & 0.5739 & $-42.6$ \\
			$360^3$ & self-affine & 0.410 & 0.5046 & 0.4515 & $-53.0$ \\
			\midrule
			$160^3$ & flat        & 0.496 & 0.8286 & 0.7699 & $-58.7$ \\
			$160^3$ & curved      & 0.738 & 0.5801 & 0.5202 & $-59.8$ \\
			$160^3$ & self-affine & 0.930 & 0.4768 & 0.4145 & $-62.3$ \\
			\bottomrule
		\end{tabular}
	\end{table}
	
	The computed $\gamma_{\mathrm{ex}}$ values of
	0.41--0.88~J/m$^2$ are comparable in magnitude to the
	0.600--0.800~J/m$^2$ range from Wolf's 0~K calculations
	summarised by Surholt and Herzig~\cite{surholt1997cu},
	and to the $0.59\pm0.12$~J/m$^2$ calorimetric estimate
	of Wegner et al.~\cite{wegner2014cu}, obtained over
	433--653~K during grain growth from 35 to 50~nm.
	Korolev et al.~\cite{korolev2022cu} further demonstrate
	the importance of boundary character and discuss temperature
	differences when comparing 0~K molecular statics with
	1273~K experiments. The comparison is limited to the overall energy scale, as
	$\gamma_{\mathrm{ex}}$ includes defect contributions beyond
	grain boundaries and uses the as-built boundary area for
	normalisation.
	
	The energy trajectories and structural observations support short-time
	stability of the tested Cu polycrystals under the 300~ps NPT protocol,
	including the smaller $(160~\text{\AA})^3$ cells. Fine-scale
	self-affine roughness nevertheless relaxes during this protocol.
	The conclusions are restricted to the tested potential, temperature,
	seed root, grain count, cell sizes, and observation time; they do not
	establish long-time microstructural stability or validate the
	interatomic potential itself. The boundary-morphology generators provide initial configurations for
	studies in which imposed GB roughness is varied systematically and its
	subsequent relaxation is measured.
	For example, the experimentally investigated Pd--Au system discussed earlier
	provides one possible target for such studies, while its reported boundary
	morphology is not used here as a validation of the present Cu models.
	More generally, the same construction framework may be applied to investigate
	how grain size, orientation statistics, boundary morphology, phase partitioning,
	and deterministic chemical decoration affect atomistic observables.
	
	% =====================================================================
	\section{Performance}
	\label{sec:performance}
	
\begin{figure}[!htb]
		\centering
		\begin{subfigure}[t]{\linewidth}
			\includegraphics[width=0.8\linewidth]{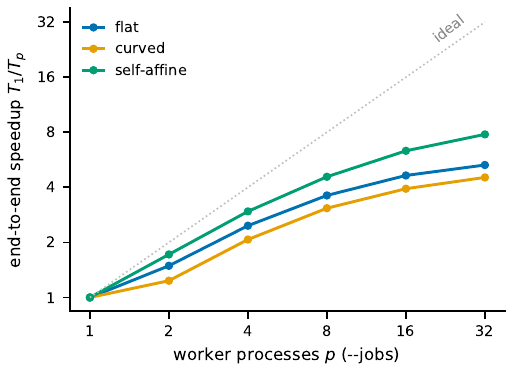}
			\caption{}\label{fig:performance-a}
		\end{subfigure}\\[2pt]
		\begin{subfigure}[t]{\linewidth}
			\includegraphics[width=0.8\linewidth]{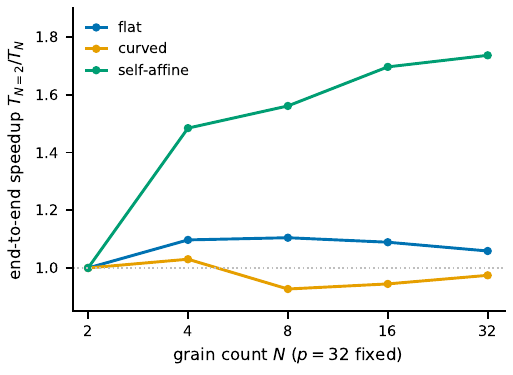}
			\caption{}\label{fig:performance-b}
		\end{subfigure}
		\caption{Parallel speedup of the three grain-boundary geometry families
			on the common $(560~\text{\AA})^3$ FCC Cu benchmark (Intel Xeon Platinum
			8480+, TRUBA HPC; complete stage-resolved timings under
			\code{benchmarks/parallel\_efficiency/}). (a)~End-to-end speedup $T_1/T_p$ versus
			worker count $p$ at 32 grains (log-log; dotted: ideal linear
			scaling). (b)~End-to-end speedup $T_{N=2}/T_N$ versus grain count
			$N$ at $p=32$ fixed.}
		\label{fig:performance}
	\end{figure}

	Performance was measured on a large-scale benchmark: the same FCC Cu
	polycrystal ($(560~\text{\AA})^3$ box, $\sim 1.5\times10^{7}$ atoms) was built with
	each of the three grain-boundary geometry families of
	Fig.~\ref{fig:geometries}(a--c), on a single Intel Xeon Platinum 8480+ node
	of the TRUBA HPC facility. Two complementary sweeps isolate the two user
	controls. Fig.~\ref{fig:performance}(a) fixes 32 grains and varies the
	worker count \code{-{}-jobs}~$\in\{1,2,4,8,16,32\}$; Fig.~\ref{fig:performance}(b)
	fixes 32 workers and varies the grain count
	$N\in\{2,4,8,16,32\}$ at constant box size. Every run completed without hard failure. The benchmark harvests stage
	timings and then discards each run's atom outputs, so the worker-count
	byte identity of Sec.~\ref{sec:determinism} is established by the
	repository's end-to-end tests rather than re-verified here; execution
	metadata are in any case excluded from that comparison. The benchmark driver and the complete
	stage-resolved records behind both panels (tessellation, fill, overlap,
	write, and remaining wall times, together with sampled peak driver-process RSS, for each
	topology and sweep point) are provided under
	\code{benchmarks/}\allowbreak\code{parallel\_efficiency/} in the repository.
	All timings were taken with the optional \code{perf} extra installed and its
	JIT cache pre-warmed, so the ownership test uses the numba route of
	Sec.~\ref{sec:parallel} throughout; the numpy fallback is slower but
	produces identical output.
	
	Fig.~\ref{fig:performance}(a) shows that the serial cost depends strongly on
	boundary geometry: $129$~s for flat, $351$~s for curved, and $896$~s for
	self-affine. The additional cost of the non-flat backends is concentrated in
	the per-candidate ownership work of the fill stage. Because this is also the
	stage most effectively parallelised by \code{-{}-jobs}, the most expensive
	geometry scales best. At 32 workers, the end-to-end speedup is $7.8\times$
	for self-affine ($896\to115$~s), $5.3\times$ for flat
	($129\to24$~s), and $4.5\times$ for curved ($351\to78$~s). Here and below,
	the quoted times are rounded to whole seconds, while the factors are
	computed from the unrounded stage-resolved timings recorded under
	\code{benchmarks/parallel\_efficiency/}. The curves flatten
	as the serial remainder dominates, as expected from Amdahl's law. For flat
	and curved geometries, that remainder is dominated by the overlap search and
	other serial stages; for self-affine geometry, the principal limiter is the
	spectral-field tessellation stage, which takes $47$~s at every worker count.
	The sampled peak resident set size (RSS) of the driver process
	ranges from $2.4$ to $3.8$~GB across the sweep, excluding worker-process memory.
	
	Fig.~\ref{fig:performance}(b) shows the second condition for parallel
	efficiency: the fill pool cannot use more workers than there are grains and
	therefore caps itself at $\min(\code{jobs},N)$, with per-grain work as the unit
	of distribution. For the fill-dominated self-affine geometry, the
	end-to-end time consequently improves by $1.7\times$ as $N$ increases from
	2 to 32 at fixed $p=32$; the fill stage itself accelerates $6.2\times$
	($145\to23$~s) as more grains become available for distribution. Flat and
	curved totals are nearly independent of $N$ because, at this box size, their
	runtimes are dominated by stages whose cost scales with atom count rather
	than grain count.
	
	For the per-grain fill stage, useful concurrency is limited
	by the grain count. Overall speedup depends on the balance
	between parallel stages and the geometry-specific serial
	remainder.
	
	% =====================================================================
	\section{Limitations and future work}
	\label{sec:conclusions}
	
	The limitations discussed in the preceding sections include
	ODF-dependent constraints on attainable $\Sigma 3$ content,
	finite-band effects on the recovered Hurst exponent, an
	orientation-dependent area overestimate of up to a factor of
	$\sqrt{3}$ for the voxel face-counting estimator, and the limited
	power of small-$N$ distribution tests. These limitations are
	documented and, where applicable, reflected in diagnostics and
	gate warnings, allowing users to assess their relevance to a
	given study.
	
	Interphase orientation relationships, including Burgers,
	Kurdjumov--Sachs, and Nishiyama--Wassermann relationships relevant
	to $\alpha/\beta$ Ti, Cu--Nb, and fcc/B2
	studies~\cite{titanium2023alphabeta,cunb_ks_bilayer,fccb2_nwks2020},
	are not generated by the current release. A planned extension is
	to combine imported multiphase fields, including lamellar
	structures supplied through \code{voxel\_import}, with
	orientation assignments consistent with these relationships.
	The current generator also does not predict equilibrium
	segregation: its shell/bulk dopant model prescribes enrichment
	geometrically. Generated configurations are starting structures
	for subsequent relaxation and physics-based simulation, rather
	than predictions of equilibrated interfaces or solute
	distributions.
	
	Future development will focus on computational efficiency and
	integrated model-building and analysis workflows. Performance
	improvements will target the bottlenecks identified in
	Sec.~\ref{sec:performance}, particularly self-affine spectral-field
	tessellation and residual overlap-search costs for flat and curved
	geometries, with additional parallelism introduced where compatible
	with the deterministic per-grain random-stream design. Complementing
	these improvements, a planned graphical studio will bring
	microstructure specification, atomistic model generation,
	visualisation, and analysis into a unified interface for researchers,
	while preserving the underlying pipeline's reproducibility and
	quantitative diagnostics.
	
	% =====================================================================
	\section{Dependencies}
	\label{sec:dependencies}
	
	\grainsmith{} requires Python 3.10 or later and depends on NumPy,
	SciPy, spglib, PyYAML, and Pydantic. Optional dependencies provide
	CIF input (ASE), DREAM.3D HDF5 import (h5py), boundary-mesh export
	(scikit-image), memory monitoring (psutil), and acceleration of
	selected geometry kernels (Numba). Installation instructions and
	optional dependency groups are documented in the user manual.
	
	% =====================================================================
	\section{Summary}
	\label{sec:impact}
	
	\grainsmith{} generates atomistic polycrystals with statistical
	control over grain sizes and volumes, crystallographic texture,
	grain-boundary disorientation angles, boundary morphology, phase
	fractions, and grain-boundary-targeted dopant enrichment.
	Within supported feature combinations, it integrates periodic
	Voronoi and SDOT-fitted Laguerre tessellations, flat, smoothly
	curved, and band-limited self-affine boundaries, and multiphase
	construction in a single pipeline. Applicable construction-time
	checks and reported structural statistics allow users to assess
	the generated configurations against their specifications.
	
	These capabilities support studies of grain-size
	effects~\cite{silicon2024grainsize,magnesium2025grainsize},
	multiphase
	microstructures~\cite{feni2024dualphase,wcco2017nanoindentation},
	and relationships among texture, boundary networks, and CSL
	character~\cite{gertsman1993texture,gbnetwork2023plasticity,NINO2023111879},
	as well as geometrically prescribed dopant enrichment.
	The generated atomic configurations and scientific data are
	byte-reproducible under the fixed-version and fixed-environment
	conditions of Sec.~\ref{sec:determinism}. Machine-readable
	provenance and reported descriptors support downstream analysis
	and quantitative comparisons across generated structures, while
	documented voxel-import workflows connect DREAM.3D and other
	voxel-based microstructure representations to atomistic simulations.

	\section*{Acknowledgements}
	The numerical calculations reported in this paper were partially performed
	at T\"UB\.ITAK ULAKB\.IM, High Performance and Grid Computing Center
	(TRUBA resources). This research received no specific grant from any funding agency in the public, commercial, or not-for-profit sectors.
	
	\section*{Data availability}
	\grainsmith{} is open-source (MIT) and available at 	\url{https://github.com/oguzhanorhan/grainsmith}. 
	
	% =====================================================================
	\bibliographystyle{elsarticle-num}
	\bibliography{cas-refs}
	
	\newpage
	% =====================================================================
	\appendix
	\setcounter{table}{0}
	\setcounter{figure}{0}
	\section{Ready-to-run example configurations}
	\label{app:examples}
	
	Every generation method discussed in Sec.~\ref{sec:theory} is accompanied by at least one
	commented YAML configuration under \cfg{examples/} in the repository, so that
	each construction discussed in this paper can be rebuilt with a single
	command. Table~\ref{tab:examples-core} catalogues every
	configuration referenced in this paper; the further \cfg{examples/advanced/}
	group, which scales six of these constructions to ${\sim}10^6$ atoms and adds a ${\sim}10^8$-atom tier in a $1500~\text{\AA}$ box, is
	catalogued in \cfg{examples/README.md}: a row's
	full repository path is its group heading followed by its file name, and the
	two subcommands of Sec.~\ref{sec:interfaces} apply unchanged to every entry,
	\code{grainsmith validate} for the schema and crystal checks (gates G1--G2)
	and \code{grainsmith generate} for the full pipeline.
	
	Successful completion requires all applicable hard checks to pass;
	warnings and diagnostic measurements are reported separately. Each file carries its own header stating what it demonstrates, its literature
	grounding where applicable, and the exact commands to run it.
	\cfg{examples/README.md} presents the same catalogue in a recommended reading
	order and records the atom count of every entry, plus the gate count for the
	\code{advanced} and \code{hpc} groups and, for \code{hpc}, the wall time
	measured on a reference run; \cfg{examples/hpc/README.md} gives the full single-node SLURM recipe. Gate
	counts differ between configurations because the registry of
	Sec.~\ref{sec:gates} is feature-conditional: a non-\code{warp} curved method
	has no G6 to evaluate, while a \code{self\_affine} spectrum adds three gates, G13 and G19 (warn) and G20 (diagnostic); the last reports a roughness index rather than a converged fractal dimension below ${\sim}5.5$ synthesis-band octaves.
	
	\begin{table*}[!t]
		\centering
		\caption{Example configurations shipped with \grainsmith{}, grouped by the
			subdirectory they live in; a row's full repository path is its group
			heading followed by its file name. The section reference on each group
			heading points to the method the group demonstrates. Atom counts are
			recorded per entry in \code{examples/README.md}, which also gives gate
			counts for the \code{advanced} and \code{hpc} groups and measured wall
			times for \code{hpc}.}
		\label{tab:examples-core}
		\footnotesize
		\setlength{\tabcolsep}{4pt}
		\begin{tabular}{@{}p{0.28\linewidth}p{0.68\linewidth}@{}}
			\toprule
			Configuration file & Demonstrates \\
			\addlinespace[1pt]
			\multicolumn{2}{@{}l}{\cfg{examples/basics/}\quad(Sec.~\ref{sec:tessellation}, Sec.~\ref{sec:multiphase})}\\[1pt]
			\cfg{b2_nial_flat.yaml} & the baseline run: ordered B2 compound, flat Voronoi boundaries, opt-in curvature analysis (flat $\Rightarrow H=K=0$ exactly) \\
			\cfg{cu_single_crystal.yaml} & \code{grains.number: 1}: the perfect-crystal baseline and gate G14 box/lattice commensurability \\
			\cfg{bicrystal_fixed_orient.yaml} & per-grain orientations from an explicit list (quaternion and axis-angle), $\Sigma5$ assignment \\
			\cfg{hcp_ti_thin_film.yaml} & hexagonal family, slab geometry with vacuum, single $2c$ Wyckoff site (ideal HCP) \\
			\addlinespace[1pt]
			\multicolumn{2}{@{}l}{\cfg{examples/crystallography/}\quad(Sec.~\ref{sec:crystal})}\\[1pt]
			\cfg{rutile_tio2_tetragonal.yaml} & tetragonal family, a free Wyckoff parameter (O $4f$), multi-sublattice stoichiometry \\
			\cfg{si_diamond_origin2.yaml} & space-group \code{setting} (origin choice 2), and why Wyckoff letters may differ from ITA there \\
			\cfg{bi_rhombohedral_R.yaml} & the rhombohedral (R) axis setting: free $(a,\alpha)$ instead of $(a,c)$ \\
			\addlinespace[1pt]
			\multicolumn{2}{@{}l}{\cfg{examples/cif/}\quad(Sec.~\ref{sec:crystal})}\\[1pt]
			\cfg{tini_cif_polycrystal.yaml} & \code{crystal.cif} direct input: monoclinic TiNi ($P2_1/m$); declared and detected group agree \\
			\cfg{ti2ni_cif_single_crystal.yaml} & CIF on the single-crystal backend: $Fd\bar{3}m$, 96 atoms/cell, three Wyckoff orbits \\
			\cfg{tini3_cif_polycrystal.yaml} & declared-vs-detected mismatch: the file declares $P1$, spglib detects $P6_3/mmc$ (warn only) \\
			\cfg{tini_cif_triclinic_single.yaml} & \code{box.cells}: triclinic single crystal as integer lattice multiples; G14 misfit below the $10^{-8}$ tolerance \\
			\cfg{tini3_cif_hexagonal_single.yaml} & \code{box.cells} on a hexagonal cell: an $xy$ tilt landing exactly on the LAMMPS bound \\
			\addlinespace[1pt]
			\multicolumn{2}{@{}l}{\cfg{examples/texture/}\quad(Sec.~\ref{sec:texture})}\\[1pt]
			\cfg{cu_twin_odf_mdf.yaml} & \code{odf\_components} ($\Sigma3$ twin pair) with \code{mdf\_target: sigma3\_angle\_enriched} annealing (G12) \\
			\cfg{fcc_cu_fiber_film.yaml} & \code{fiber} texture ($\langle111\rangle\parallel z$ with spread), slab with vacuum, per-atom GB margin \\
			\addlinespace[1pt]
			\multicolumn{2}{@{}l}{\cfg{examples/grain_geometry/}\quad(Sec.~\ref{sec:tessellation}, Sec.~\ref{sec:curved})}\\[1pt]
			\cfg{cu_lognormal_sizes.yaml} & log-normal grain volumes via SDOT-fitted Laguerre weights (G11), plus an EBSD-like section \\
			\cfg{cu_equal_sizes.yaml} & \code{size\_distribution.type: equal}: every grain fitted to $V_{\mathrm{box}}/N$ \\
			\cfg{cu_volume_list_sizes.yaml} & \code{type: volumes}: an explicit relative per-grain volume list (bimodal $3{:}1$) \\
			\cfg{al_equiaxed_lloyd_weights.yaml} & Lloyd centroidal relaxation with \code{additive\_weights} curved boundaries, and the opt-in PLY boundary mesh (\code{output.mesh}) \\
			\cfg{fe_bcc_anisotropic.yaml} & \code{anisotropic} (ellipsoidal-metric) elongated grains with the \code{midpoint\_merge} policy \\
			\cfg{cu_curved_sizes.yaml} & the one legal curved-plus-prescribed-volume combination: \code{warp} over a \code{power} base \\
			\cfg{fcc_cu_curved_warp.yaml} & large scale: domain-warp curved boundaries at $(480~\text{\AA})^3$, G6 guard, curvature analysis \\
			\addlinespace[1pt]
			\multicolumn{2}{@{}l}{\cfg{examples/alloys_multiphase/}\quad(Sec.~\ref{sec:multiphase})}\\[1pt]
			\cfg{ti_alpha_beta.yaml} & \code{phases:} HCP $\alpha$ + BCC $\beta$ Ti at 60/40 volume fractions, interphase rows (G15) \\
			\cfg{cufe_composite.yaml} & phases differing in element: FCC Cu + BCC Fe, two LAMMPS atom types \\
			\cfg{cuni_solid_solution.yaml} & stochastic site occupancy (\code{Cu: 0.9, Ni: 0.1}) and gate G9 composition drift \\
			\addlinespace[1pt]
			\multicolumn{2}{@{}l}{\cfg{examples/doping/}\quad(Sec.~\ref{sec:doping})}\\[1pt]
			\cfg{al_mg_gb_substitutional.yaml} & substitutional mode: 2~at\% Mg replacing Al in place, GB-targeted dopant placement $E=5$ \\
			\cfg{fe_c_gb_interstitial.yaml} & interstitial mode: 2~at\% C on the \code{bcc\_octahedral} preset sublattice (G17, G18) \\
			\cfg{tio2_cif_li_interstitial.yaml} & CIF-sourced interstitial sublattice: Li in anatase by symmetry-orbit expansion \\
			\addlinespace[1pt]
			\multicolumn{2}{@{}l}{\cfg{examples/self\_affine\_gb/}\quad(Sec.~\ref{sec:curved})}\\[1pt]
			\cfg{pdau_perturbed_self_affine.yaml} & \code{perturbed\_distance} with \code{spectrum: self\_affine} on Pd--Au (G13, G19, G20) \\
			\cfg{tio_2_thin_film_self_affine.yaml} & self-affine boundaries in a slab box (periodic $xy$, free $z$ with vacuum) \\
			\addlinespace[1pt]
			\multicolumn{2}{@{}l}{\cfg{examples/voxel\_import/}\quad(Sec.~\ref{sec:curved})}\\[1pt]
			\cfg{fe_voxel_import.yaml} & atomise an external voxel label field (shipped $32^3$ \code{.npy}; DREAM.3D via an extra) \\
			\addlinespace[1pt]
			\multicolumn{2}{@{}l}{\cfg{examples/hpc/}\quad(Sec.~\ref{sec:parallel}, Sec.~\ref{sec:performance})}\\[1pt]
			\cfg{1_cu_flat.yaml} & flat Voronoi boundaries at cluster scale; the single-node \code{sbatch}/\code{salloc} recipe \\
			\cfg{2_cu_curved.yaml} & \code{anisotropic} elongated grains ($\le2.5{:}1$) with \code{midpoint\_merge} at cluster scale \\
			\cfg{3_cu_fractal.yaml} & \code{perturbed\_distance} with \code{spectrum: self\_affine} ($\mathcal{H}=0.8$) at cluster scale \\
			\bottomrule
		\end{tabular}
	\end{table*}
	
	% =====================================================================
	% APPENDIX B - the construction-time QA-gate registry.
	\section{The construction-time QA-gate registry}
	\label{app:gates}
	
	Table~\ref{tab:gates} lists the full registry referenced throughout
	Sec.~\ref{sec:gates} and Sec.~\ref{sec:validation}. A \emph{hard} gate aborts
	the run, a \emph{warn} gate completes the run and records a flag in
	\code{summary.csv}, and a \emph{diagnostic} gate never trips: it records a quantity that has no defensible pass/fail threshold. G20, G23, G25 and G26 are diagnostics in this sense and never generate a warning row by themselves. Every named threshold is
	defined in a single constants module; a name in \code{typewriter} type is the
	constant or configuration field that carries it, while the fixed percentage
	bands of G8, G9 and G17 are given numerically.
	
	\begin{table*}[!p]
		\centering
		\caption{The construction-time QA-gate registry. ``hard'' aborts the run; ``warn'' records a flag and continues; ``diagnostic'' never trips and only reports a quantity for interpretation. $^{\dagger}$G5 aborts the run for the tessellation backends; \code{voxel\_import} demotes it to a warning, and under \code{perturbed\_distance} repaired components are reported through G19 instead. G22--G26 cover the
			volume-weighted ODF and the angle-window versus true-CSL $\Sigma3$ gap
			(Sec.~\ref{sec:texture}).}
		\label{tab:gates}
		\footnotesize
		\setlength{\tabcolsep}{5pt}
		\begin{tabular}{@{}cp{0.34\linewidth}p{0.34\linewidth}l@{}}
			\toprule
			Gate & Check & Threshold (constant/field) & Kind \\
			\midrule
			G1  & config schema + 29 cross-field rules & pydantic \code{extra='forbid'} + resolver & hard \\
			G2  & spglib symmetry round-trip of the reference unit cell
			& ITA number must match the reference group: requested for Wyckoff input, initially detected for CIF input
			& hard \\			
			G3  & tessellation volume sum = box         & flat \code{VOL\_REL\_TOL=1e-6} & hard \\
			G4  & per-cell Euler characteristic $N_V-N_E+N_F=2$ & exact (combinatorial) & hard \\
			G5  & grain connectivity / no empty cells   & \code{strict\_connectivity}; per-method severity, see caption & hard$^{\dagger}$ \\
			G6  & warp bijectivity guards               & \code{WARP\_GRAD\_MAX=0.5} & hard \\
			G7  & post-overlap minimum distance         & overlap \code{cutoff} ($0.85\,d_{nn}$) & hard \\
			G8  & final atom count vs theoretical       & warn $>2\%$, fail $>5\%$ & warn/hard \\
			G9  & composition drift vs stoichiometry    & warn $>1\%$ per species & warn \\
			G10 & LAMMPS file re-parses to same system  & exact count + bounds & hard \\
			G11 & re-measured grain-volume error        & \code{vol\_tol} (def.\ 1e-3) & hard \\
			G12 &  disorientation-angle-target $\chi^2$ after annealing  & \code{chi2\_max} (def.\ 0.5) & warn \\
			G13 & back-estimated Hurst vs target        & \code{HURST\_G13\_TOL=0.15} & warn \\
			G14 & single-crystal commensurability       & \code{COMMENSURATE\_TOL=1e-8} & warn \\
			G15 & re-measured phase volume fractions
			& granularity scale $V_{\max}/V_{\mathrm{box}}$ & warn \\			
			G16 & GB-curvature degenerate-sample fraction & \code{CURV\_G16\_DROP\_TOL=0.05} & warn \\
			G17 & dopant composition + enrichment       & drift $>1$\,pp; enrichment $>15\%$ rel. & warn \\
			G18 & post-placement dopant min distance    & per-dopant \code{min\_distance} (fresh query) & hard \\
			G19 & \code{perturbed\_distance} G5-repair reassigned-voxel fraction & \code{REASSIGNED\_FRACTION\_WARN\_TOL=0.02} & warn \\
			G20 & section-perimeter box-counting roughness index $D_b$ (\code{perturbed\_distance})
			& fit restricted to $[l_{\min},l_{\max}]$; no pass/fail threshold; unavailable if no usable fit
			& diagnostic \\
			G21 & magnitude of the per-grain face-interior Gaussian-curvature integral, Eq.~\eqref{eq:g21}
			& \code{CURV\_G21\_GAUSS\_BONNET\_TOL}$=4\pi$; heuristic reference scale
			& warn \\
			G22 & discrete volume-weighted orientation drift; optional kernel MMD diagnostic
			& drift $>\code{odf\_drift\_max}+10^{-9}$ if capped; MMD outside the central 95\% null interval if evaluated
			& warn \\
			G23 & $\Sigma3$ angular-window area fraction vs true CSL $\Sigma3$ area fraction & none: diagnostic pair, single-phase runs only      & diagnostic \\
			G24 & final (post-warp) per-grain volume vs target & \code{WARP\_VOLUME\_G24\_TOL}$=0.10$ & warn \\
			G25 & configured texture-component weight vs realised count and volume fractions & none: \code{odf\_components} scheme only 
			& diagnostic \\
			G26 & tessellation-volume vs atom-count grain-weight discrepancy, Eq.~\eqref{eq:weighting-gap}
			& none: reports total variation, maximum weight difference, and mean atoms per grain; evaluated within each phase
			& diagnostic \\
			\bottomrule
		\end{tabular}
	\end{table*}
	
\end{document}